\documentclass[aps,prb,twocolumn,unsortedaddress,showpacs,preprintnumbers]{revtex4}

\usepackage[pdftex]{graphicx}

\begin{document}


\title{Unified Electronic Phase Diagram for Hole-Doped High-$T_c$ Cuprates}

\author{T. Honma}
\affiliation{Department of Physics, Asahikawa Medical College, Asahikawa, Hokkaido 078-8510, Japan}
\email{honma@asahikawa-med.ac.jp}

\author{P. H. Hor}
\affiliation{Department of Physics and Texas Center for Superconductivity, University of Houston, Houston, TX77204-5005, U.S.A.}
\email{phor@uh.edu}


\author{}


\date{\today}

\begin{abstract}
We have analyzed various characteristic temperatures and energies of hole-doped high-$T_c$ cuprates as a function of a dimensionless hole-doping concentration ($p_u$). Entirely based on the experimental grounds we construct a unified electronic phase diagram (UEPD), where three characteristic temperatures ($T^*$'s) and their corresponding energies ($E^*$'s) converge as $p_u$ increases in the underdoped regime. $T^*$'s and $E^*$'s merge together with the $T_c$ curve and 3.5$k_BT_c$ curve at $p_u$ $\sim$ 1.1 in the overdoped regime, respectively. They finally go to zero at $p_u$ $\sim$ 1.3. The UEPD follows an asymmetric half-dome-shaped $T_c$ curve in which $T_c$ appears at $p_u$ $\sim$ 0.4, reaches a maximum at $p_u$ $\sim$ 1, and rapidly goes to zero at $p_u$ $\sim$ 1.3. The asymmetric  half-dome-shaped $T_c$ curve is at odds with the well-known symmetric superconducting dome for La$_{2-x}$Sr$_x$CuO$_4$ (SrD-La214), in which two characteristic temperatures and energies converge as $p_u$ increases and merge together at $p_u$ $\sim$ 1.6, where $T_c$ goes to zero. The UEPD clearly shows that pseudogap phase precedes and coexists with high temperature superconductivity in the underdoped and overdoped regimes, respectively. It is also clearly seen that the upper limit of high-$T_c$ cuprate physics ends at a hole concentration that equals to 1.3 times the optimal doping concentration for almost all high-$T_c$ cuprate materials, and 1.6 times the optimal doping concentration for the SrD-La214. Our analysis strongly suggests that pseudogap is a precursor of high-$T_c$ superconductivity, the observed quantum critical point inside the superconducting dome may be related to the end point of UEPD, and the normal state of the underdoped and overdoped high temperature superconductors cannot be regarded as a conventional Fermi liquid phase. 
\end{abstract}


\maketitle

\section{Introduction}

The unique hallmark of high temperature superconductors (HTSs) is a pseudogap phase characterized by the observation of a multiple pseudogap temperatures ($T^*$'s) and pseudogap energies ($E^*$'s) by a large number of different experimental probes. While the pseudogap phase precedes the high temperature superconducting phase characterized by the superconducting transition temperature ($T_c$) and superconducting gap energy ($\Delta_c$), it is not clear how $T^*$, $T_c$, $E^*$, and $\Delta_c$ are related to each other. Specifically, how are $T^*$ and $E^*$ related to the occurrence of the high-$T_c$ superconductivity is still unclear. Is pseudogap a sufficient and/or necessary condition for high $T_c$ or is it just a complication of specific material systems? Is it collaborating or competing with superconductivity? For instance, it is argued that the pseudogap is a competing order that may have nothing to do with high $T_c$.\cite{tal01} On the other hand it is also suggested that the pseudogap is intimately related to high $T_c$.\cite{bat96,eme97} To distinguish these two contradictory pictures that are critical to the mechanism of high-$T_c$ superconductivity requires a comparison of various characteristic temperatures and energies in a universal phase diagram for all HTSs. Any systematic behavior derived from this kind of phase diagram will provide true intrinsic properties of HTS that are free from material-specific complications. However, up until now there is no such a comparison made and no such phase diagram available. We have analyzed numerous published data in the literature. We carefully select 27 HTSs: 11 single-layer, 11 double-layer and five triple-layer HTSs as summarized in the Table~\ref{tab:table1}. The selection criteria will follow when we discuss the construction of the figures. There are 15 different experimental probes used for these 27 HTSs which are summarized in Table~\ref{tab:table2}. In this paper we unify the characteristic temperatures of all these data of 27 HTSs on one single phase diagram entirely based on our proposed universal hole concentration scale that itself is also based on experimental results.

In the single-layer SrD-La214, where the hole-doping concentration can be unambiguously determined from the Sr-content ($x$),\cite{tor88} $T_c$($x$) exhibits a well-known symmetric bell-shaped curve, i.e., the so-called superconducting dome, with a maximum $T_c$ ($T_c^{max}$) located at $x$ $\sim$ 0.16.\cite{pre91} The symmetrical dome-shaped $T_c$ curve or the superconducting dome is approximately represented by the following parabola, 
\begin{eqnarray}
 1 - \frac{T_c}{T_c^{max}} =
 \left.
 \begin{array}{ll}
 82.6  \times (x-0.16)^2.  & \label{form1}
\end{array}
\right.
\end{eqnarray}
Assuming that all HTSs have the identical symmetric superconducting dome, $x$ can be replaced with the hole-doping concentration ($P_{T_c}$). Then, this relation could be used to determine the hole-doping concentration for many other HTSs.\cite{pre91,naq05,obe92,ber96,wat00,sut03,haw07,tak01,cam99,din01,tan06,miy98,ozy00,moo93,blu97,cap07,gas97,ken95,sug03} Using this hole-scale based on the superconducting dome, the $P_{T_c}$-scale, various phase diagrams have been constructed.\cite{tal01} A distinct feature in one of such phase diagrams is that $T^*$ crosses the superconducting dome and reaches zero at a quantum critical point (QCP) inside the dome.\cite{tal01,naq05} On the other hand, without using the $P_{T_c}$ scale, some qualitative experimental observations seem to support another picture, where $T^*$ touches the superconducting dome at around $T_c^{max}$ and merges into the superconducting dome with no QCP inside the dome.\cite{bat96} To distinguish these two fundamentally different pictures, we need a hole scale that can reveal the true intrinsic doping dependences of $T^*$, $T_c$, $E^*$, and $\Delta_c$, which have been already observed.

\begin{table}[b]
\caption{\label{tab:table1}The chemical formula and the notation for the HTSs used in the present work.}
\begin{ruledtabular}
\begin{tabular}{ll}
 Chemical formula & Notation \\ \hline
 (Single-layer HTS) \\
 La$_{2-x}$Sr$_x$CuO$_4$ & SrD-La214 \\
 La$_{2-x}$Ba$_x$CuO$_4$ & BaD-La214 \\
 La$_2$CuO$_4$ & OD-La214 \\ 
 (Nd$_{1.6-x}$Ce$_x$Sr$_{0.4}$)CuO$_4$ & CeD-NdSr214 \\
 (La$_{1.6-x}$Nd$_{0.4}$Sr$_x$)CuO$_4$ & SrD-LaNd214 \\
 Tl$_2$Ba$_2$CuO$_{6+\delta}$ & OD-Tl2201 \\
 Bi$_2$Sr$_{2-x}$La$_x$CuO$_{6+\delta}$ & CD-Bi2201 \\
 (Bi$_{1.74}$Pb$_{0.38}$)Sr$_{1.88}$CuO$_{6+\delta}$ & OD-BiPb2201 \\
 (Bi$_{1.35}$Pb$_{0.85}$)(Sr$_{1.47-x}$La$_{0.38+x}$)CuO$_{6+\delta}$ & CD-BiPb2201 \\
 HgBa$_2$CuO$_{4+\delta}$ & OD-Hg1201 \\
 Tl$_{1-x}$Pb$_x$Sr$_2$CuO$_{5-\delta}$ & CD-TlPb1201 \\
\hline  
 (Double-layer HTS) \\           
 Y$_{1-x}$Ca$_x$Ba$_2$Cu$_3$O$_6$ & CaD-Y1236 \\
 YBa$_2$Cu$_3$O$_{6+\delta}$ & OD-Y123 \\
 Y$_{1-x}$Ca$_x$Ba$_2$Cu$_3$O$_{6+\delta}$ & CD-YCa123 \\
 (Ca$_{1-x}$La$_{x}$)(Ba$_{1.75-x}$La$_{0.25+x}$)Cu$_3$O$_{6+\delta}$ & CLBLCO \\
 CaLaBaCu$_3$O$_{6+\delta}$ & CLBCO \\ 
 Bi$_2$Sr$_2$CaCu$_2$O$_{8+\delta}$ & OD-Bi2212 \\
 Bi$_2$Sr$_2$(Ca$_{1-x}$Y$_x$)Cu$_2$O$_{8+\delta}$ & CD-Bi2212 \\
 HgBa$_2$CaCu$_2$O$_{6+\delta}$ & OD-Hg1212 \\
 (Hg$_{0.5}$Fe$_{0.5}$)Ba$_2$(Ca$_{1-x}$Y$_x$)Cu$_2$O$_{6+\delta}$ & CD-HgFe1212\\
 Tl(BaSr)CaCu$_2$O$_{6+\delta}$ & CD-Tl1212 \\
 (Tl$_{0.5+x}$Pb$_{0.5-x}$)Sr$_2$(Ca$_{1-y}$Y$_y$)Cu$_2$O$_{6+\delta}$ & CD-TlPb1212 \\
 \hline
  (Triple-layer HTS) \\
 Bi$_2$Sr$_2$CaCu$_3$O$_{10+\delta}$ & OD-Bi2223 \\
 HgBa$_2$Ca$_2$Cu$_3$O$_{8+\delta}$ & OD-Hg1223 \\
 TlBa$_2$Ca$_2$Cu$_3$O$_{8+\delta}$ & OD-Tl1223 \\
 (Cu$_{1-x}$Ca$_x$)Ba$_2$Ca$_2$Cu$_3$O$_{8+\delta}$ & CD-CuCa1223 \\
 (Cu$_{1-x}$C$_x$)Ba$_2$Ca$_2$Cu$_3$O$_{8+\delta}$ & CD-CuC1223 \\             
\end{tabular}
\end{ruledtabular}
\end{table}

\begin{table}[b]
\caption{\label{tab:table2}The experimental probes and their notations for the present work.}
\begin{ruledtabular}
\begin{tabular}{ll}
 Experimental probe & Notation \\ \hline
 resistivity & $\rho$ \\
 $a$-axis resistivity & $\rho_a$ \\
 $c$-axis resistivity & $\rho_c$ \\
 in-plane resistivity & $\rho_{ab}$ \\
 inflection point of $\rho$($T$) & $d^2$$\rho$/$dT^2$ \\
 thermoelectric power & TEP or $S$ \\
 TEP at 290 K & $S^{290}$ \\
 $a$-axis TEP & $S_a$ \\
 in-plane TEP & $S_{ab}$ \\
 susceptivility & $\chi$ \\ 
 susceptivility (H//$c$) & $\chi_c$ \\  
 susceptivility (H//$ab$) & $\chi_{ab}$ \\ 
 nuclear magnetic resonance  & NMR \\
 nuclear quardruple resonance & NQR \\  
 spin-lattice relaxation rate (NQR) & $(T_1T)^{-1}$ \\
 NMR knight shift (H//$c$) & $K_c$ \\ 
 angle-resolved photoemission spectroscopy & ARPES \\ 
 angle-integrated photoemission spectroscopy & AIPES \\
 superconductor-insulator-superconductor \\ tunneling & SIS \\
 superconductor-insulator-normal metal \\ tunneling & SIN \\ 
 near edge x-ray absorption fine structure & NEXAFS \\
 electronic specific heat coefficient & $\gamma$ \\
 thermal conductivity & $\kappa$ \\ 
 neutron scattering & $neutron$ \\ 
 electronic Raman scattering & $ERS$ \\ 
 quasiparticle relaxation rate & $QPR$ \\ 
 polar angular magnetoresistance oscillations & $AMRO$ \\ 
\end{tabular}
\end{ruledtabular}
\end{table}

\begin{figure}
\includegraphics[scale=0.5]{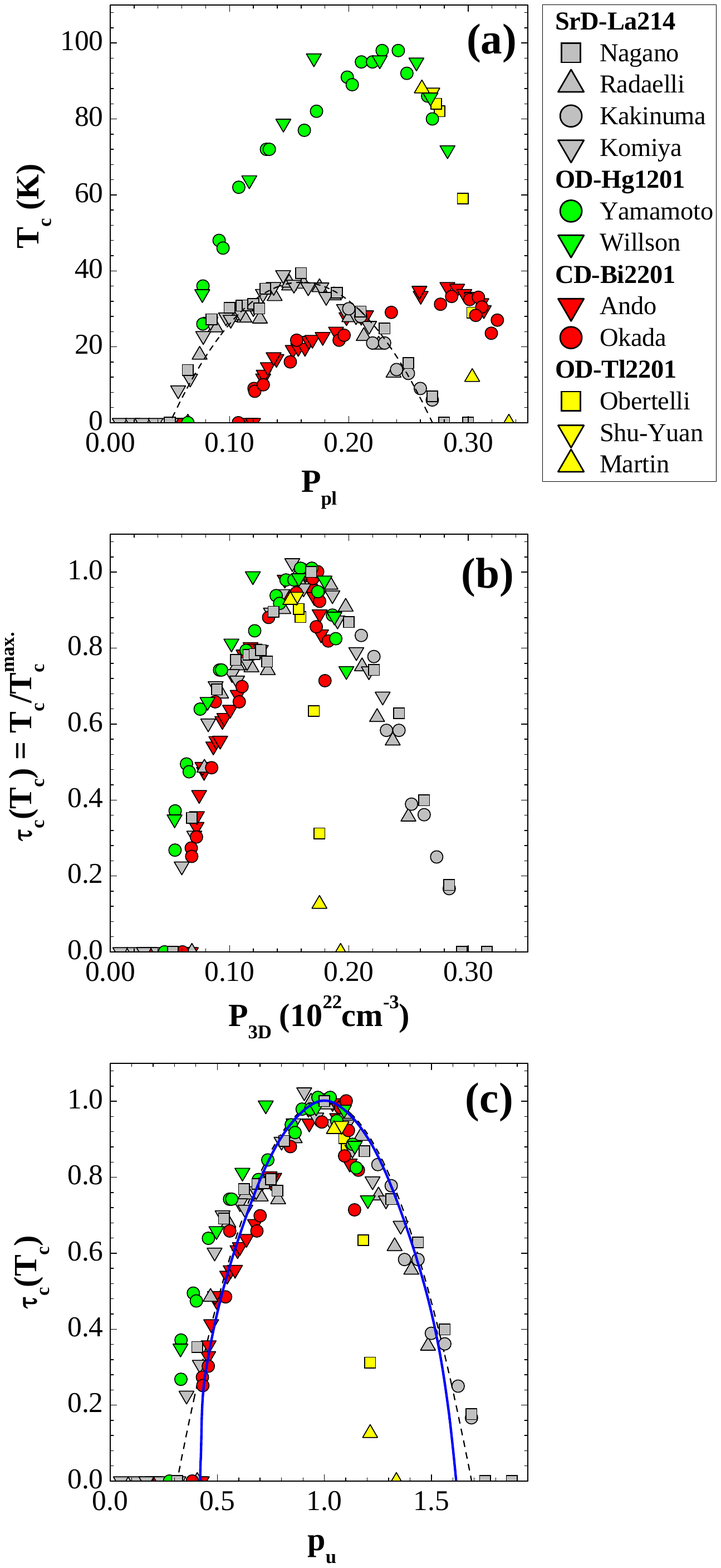}
\caption{\label{fig1} (Color online) For the single-layer HTSs, (a) the superconducting transition temperature ($T_c$) as a function of $P_{pl}$, (b) the reduced superconducting transition temperature $\tau_c$($T_c$) ($\equiv$ $T_c$/$T_c^{max}$) as a function of the effective 3D hole-doping concentration $P_{3D}$, and (c) the $\tau_c$($T_c$) as a function of $P_{3D}$. The plotted data are summarized in the Table~\ref{tab:table3}. The broken line comes from the equation (\ref{form1}). The solid line is our superconducting dome.}
\end{figure}

The common structural features of HTS are CuO$_2$-planes that host the doped holes and the block layers that supply the holes into the planes through oxygen-doping and/or cation-doping. While the doped hole-carriers are initially confined in the CuO$_2$-planes sandwiched between the insulator-like block layers, the holes are partially deconfined from the planes with doping. Therefore, the lightly doped HTS generally shows strongly two-dimensional (2D) properties. However, as the hole-doping increases, some physical properties are 2D and some, although built on the 2D carriers, will nominally be three-dimensional (3D) in nature. Therefore, it is necessary to use 2D and 3D carrier-doping concentrations to address 2D and 3D physical properties, respectively. To quantitatively study such dimensionality-dependent physical properties we have proposed a universal \textit{\textbf{planar}} hole scale ($P_{pl}$-scale) for determining the hole-doping content per CuO$_2$ plane ($P_{pl}$).\cite{hon04} In this scale, the $P_{pl}$ is uniquely determined from  $S^{290}$.\cite{hon04} We showed that, in Ref.\ \onlinecite{hon04}, the $P_{pl}$-scale is independent of the nature of the dopant, the number of CuO$_2$-plane layers per formula unit cell ($n_{layer}$), the structure and the sample quality, namely, single crystal or not. This universal $S^{290}$($P_{pl}$)-relationship is built on the sound experimental observations, which is similar to the situation of the most popular $P_{T_c}$-scale, although it is still mainly empirical and waited to be theoretically justified. Since the average area per copper in the CuO$_2$-plane is almost independent of the HTS materials, therefore, $P_{pl}$ is essentially equal to 2D hole-doping concentration defined as the hole-doping content per unit area. Using the 2D $P_{pl}$-scale, it was found in the phase diagram for all major HTSs plotted as a function of $P_{pl}$ that the $T^*$-curves are independent of the $n_{layer}$ while the $T_c$-curve strongly depends on it.\cite{hon04} Therefore the $P_{pl}$-scale is intrinsically consistent with the pseudogap energy scale.\cite{hon04} We can also extend the hole-doping content per CuO$_2$ plane to an effective 3D hole-doping content per CuO$_2$ block, which includes the oxygen coordination around the plane, ($P_{3D}$) by a simple conversion formula $P_{3D}$ $\equiv$ $P_{pl}$ $\times$ ($n_{layer}$/$V_{u.c.}$), where $V_{u.c.}$ is the unit cell volume.\cite{hon06} Since $P_{3D}$ is essentially the hole-doping content per unit volume, therefore, this natural extension of $P_{pl}$-scale to $P_{3D}$ ($P_{3D}$-scale) has allowed us to address the corresponding 3D properties. \cite{hon06} For instance, in the case of the single-layer HTS, the Hall number per ``$cm^3$'', calculated from the in-plane Hall coefficient, is not scaled with $P_{pl}$ but $P_{3D}$.\cite{hon06} The $\tau_c$($T_c$), a reduced temperature-scale defined as $\tau_c$($T$) $\equiv$ $T$/$T_c^{max.}$, of the single-layer HTS universally appears at 6 $\times$ 10$^{20}$ cm$^{-3}$ and reaches the $T_c^{max.}$ at 1.6 $\times$ 10$^{21}$ cm$^{-3}$ as shown in Fig.\ \ref{fig1}(b),\cite{hon06} although their critical hole-doping concentrations on the $P_{pl}$-scale depend on the materials as shown in Fig.\ \ref{fig1}(a).\cite{hon04} Thus, it was shown that various normal and superconducting properties for many different material systems can be consistently compared by using either $P_{pl}$ or $P_{3D}$.\cite{hon04,hon06,hon07}

\begin{table}
\caption{\label{tab:table3}The $T_c^{max.}$ and $P_{pl}^{opt.}$ for single-layer HTSs plotted in Figs.\ \ref{fig1}(a) -\ \ref{fig1}(c).}
\begin{ruledtabular}
\begin{tabular}{lcll}
 \qquad HTS & $T_c^{max.}$ (K) & $P_{pl}^{opt.}$ & Ref(s). \quad \\ \hline
 \quad  SrD-La214 & 39.4 & 0.16 & \onlinecite{nag93} \\
 \quad  SrD-La214 & 37 & 0.16 & \onlinecite{rad94} \\ 
 \quad  SrD-La214 & 36 & 0.16 & \onlinecite{kak99} \\
 \quad  SrD-La214 & 38 & 0.16 & \onlinecite{kom05} \\    
 \quad  OD-Hg1201 & 97 & 0.235 & \onlinecite{yam00,wil00} \\ 
 \quad  CD-Bi2201 & 35.5  & 0.28 & \onlinecite{and00} \\
 \quad  CD-Bi2201 & 33  & 0.28 & \onlinecite{oka06} \\
 \quad  OD-Tl2201 & 93$^{\text{a}}$  & 0.25$^{\text{a}}$ & \onlinecite{obe92,shu93,mar95} \\                 
\end{tabular}
\end{ruledtabular}
\footnotetext[1]{We use the reported highest $T_c$ = 93 K as $T_c^{max.}$ (Ref.\ \onlinecite{wag97}). From the plot of $T_c$ vs $P_{pl}$ in Fig.\ \ref{fig1}(a), the optimal $P_{pl}$ is estimated to be $\sim$0.25. The detail is in the text.}
\end{table}

In order to reveal the intrinsic generic electronic properties of all HTSs, it is necessary to be able to put both 2D and 3D physical properties on a single phase diagram. To achieve this goal, we need a carrier-scale that is not only independent of the material system but also independent of the dimensionality of the physical properties. This can be achieved if, for each material system, we scale $P_{pl}$ and $P_{3D}$ with their corresponding optimal doping concentrations, $P_{pl}^{opt.}$ and $P_{3D}^{opt.}$, respectively. Here, we introduce a dimensionless unified hole-doping concentration, $p_u$ ($p_u$ $\equiv$ $P_{pl}$/$P_{pl}^{opt.}$ = $P_{3D}$/$P_{3D}^{opt.}$). This unified hole scale ($p_u$ scale) can be used for all physical properties, which is independent of their dimensionality, in all HTSs. Indeed, the identical doping dependent behaviors are preserved even though $\tau_c$($T_c$) of the single-layer HTS plotted as a function of $P_{3D}$ in Fig.\ \ref{fig1}(b) was replotted as a function of $p_u$ in Fig.\ \ref{fig1}(c).\cite{rad94,nag93,kom05,yam00,wil00,and00,oka06,obe92,shu93,mar95,kak99} Here, each $P_{pl}^{opt.}$ was determined from the plot of $T_c$ vs $P_{pl}$ for the each compound in the present work or Refs.\ \onlinecite{hon04,hon06,hon07}. For the OD-Tl2201 there was few reports on the optimally doped samples because the optimally doped OD-Tl2201 is hard to prepare. In this case we use the highest $T_c$ = 93 K among the published data as $T_c^{max.}$.\cite{wag97} From the plot of $T_c$ vs $P_{pl}$ in Fig.\ \ref{fig1}(a), the optimal $P_{pl}$ is estimated to be $\sim$0.25. They are summarized in Table~\ref{tab:table3}. Essentially we can view $p_u$ as a scaled dimensionality- and material- independent universal carrier-doping concentration that preserves the intrinsic doping dependency for \textit{\textbf{any}} physical property for \textit{\textbf{all}} HTSs. In this paper, we have analyzed the characteristic temperatures and energies observed in the 27 HTSs by 15 different experimental probe as a function of $p_u$. We find a dopant-specific unified electronic phase diagram for HTS. The dominate phase diagram is an asymmetric half-dome-shaped $T_c$-curve for the cation and anion (oxygen) co-doped (CD) HTS. $T_c$ for the purely oxygen-doped (OD) HTS also follows the half-dome-shaped $T_c$ curve with some indication of the influence of the thermally induced oxygen redistribution.

\section{Analysis}

\begin{figure*}
\includegraphics[scale=0.5]{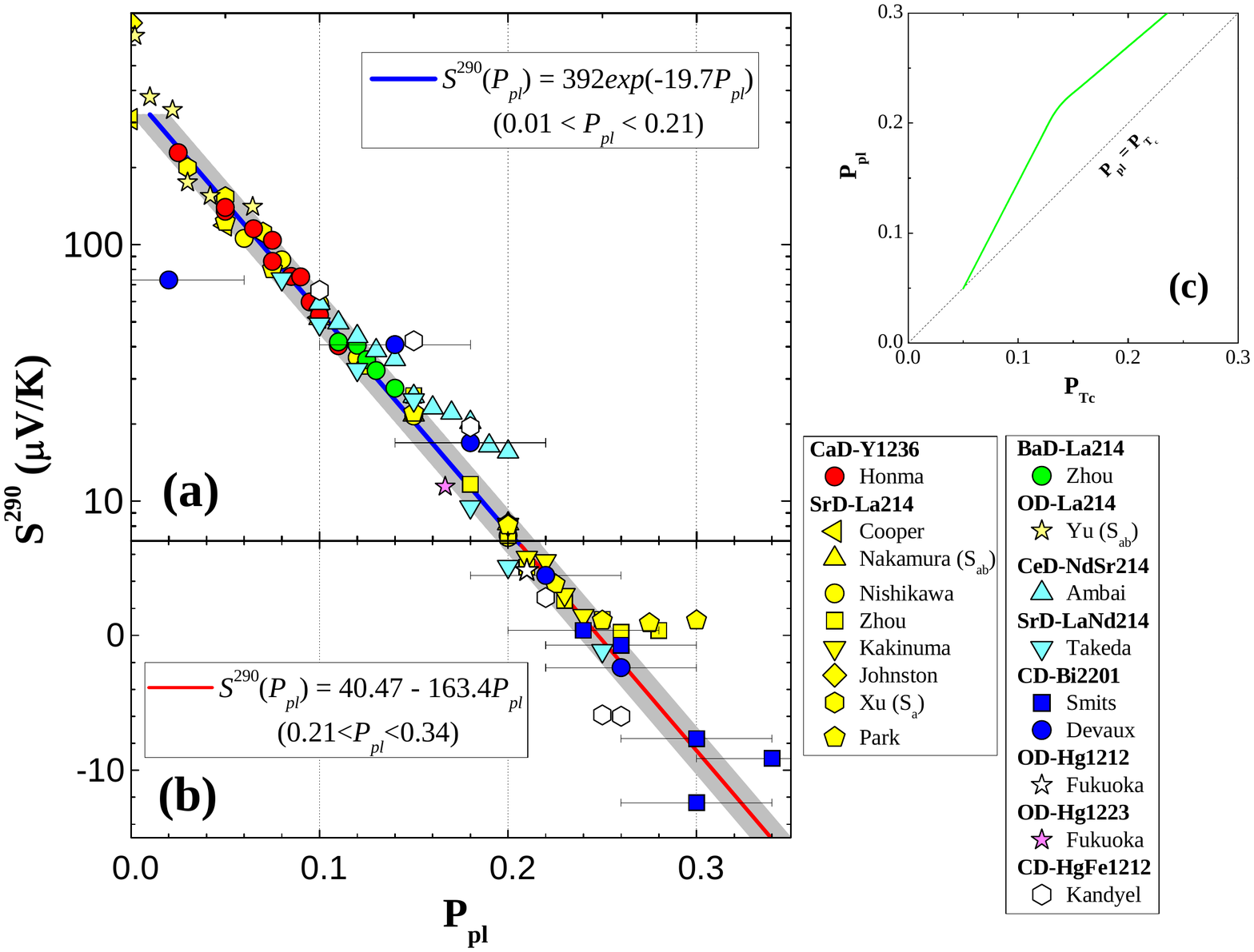}
\caption{\label{fig2} (Color online) $S^{290}$ as a function of the hole-doping content per CuO$_2$ plane. (a) $S^{290}$ ($\geq$ 7 $\mu$V/K) on the upper panel is plotted on a logarismic scale, while (b) $S^{290}$ ($<$ 7 $\mu$V/K) on the lower panel is plotted on a linear scale. The plotted data are summarized in the Table~\ref{tab:table4}. (c) Quantitative comparison between $P_{pl}$ and $P_{T_c}$. The dotted line shows $P_{pl}$ = $P_{T_c}$. We used this relation for the conversion from $P_{T_c}$ into $P_{pl}$. The error of $P_{pl}$ is below 0.04 for the CD-Bi2201 and below 0.01 for all other HTSs. The error bar for the other materials is not shown. The shaded area represents a region with the $P_{pl}$-error of $\pm$0.01 around the universal $S^{290}$($P_{pl}$)-curve.}
\end{figure*}

The details of how the $P_{pl}$- and $P_{3D}$-scales were constructed had been reported in Refs.\ \onlinecite{hon04} and \onlinecite{hon06}, respectively. The determination of $P_{pl}$ based on TEP is most reliable. Accordingly, the data including TEP are selected among the accumulated published data. The second reliable determination of $P_{pl}$ is determined from the value of $T_c$ using $T_c$ vs $P_{pl}$ curve for each compound reported in Refs.\ \onlinecite{hon04,hon06,hon07}. When the data with $P_{T_c}$ is analyzed, as the third method of determining $P_{pl}$, the $P_{pl}$ is converted from $P_{T_c}$ by using the relation in Fig.\ \ref{fig2}(c) discussed below. To clearly label how $P_{pl}$ was determined for each sample or data set used in this paper, we use the following character to designate such that I to be the second method if the cited data have no TEP but $T_c$ and II to be the third method if the cited data has only $P_{T_c}$. This designation to indicate the origin of the $P_{pl}$ will be used in Table~\ref{tab:table5} - ~\ref{tab:table9} and in Figs.\ \ref{fig3} - \ \ref{fig6}. We will use no designation whenever $P_{pl}$ is directly determined from the TEP. All the HTSs used in the present analysis are summarized in Table~\ref{tab:table1}.

We examine various characteristic temperatures and energies of HTSs for constructing the phase diagram. The pseudogap is generally observed as the characteristic temperature derived by a scaling of the temperature dependence, as a distinct change in the slope of the temperature dependence or as a peak value in the energy-dispersion at a fixed temperature. Therefore, a reliable estimation can only be achieved through using a wide temperature or energy range. We only chose the characteristic temperatures and energies obtained by direct observation or those obtained through careful analysis of the data covering a wide temperature or energy range. For example, when $T^*$ is derived by the scaling of the temperature dependence observed below 300 K, $T^*$'s over 300 K is not used.

\begin{table}
\caption{\label{tab:table4}The HTSs plotted in Figs.\ \ref{fig2}(a) and\ \ref{fig2}(b).}
\begin{ruledtabular}
\begin{tabular}{clcl}
 \quad $n_{layer}$ & \quad  HTS &  TEP &  Ref(s). \\ \hline
 1 & SrD-La214 & $S$ & \onlinecite{coo87,nis94,zho95,kak99,joh87,par05} \quad \\
 & SrD-La214 & $S_{ab}$ & \onlinecite{nak93} \\
 & SrD-La214 & $S_{a}$ & \onlinecite{xu00} \\
 & BaD-La214 & $S$ & \onlinecite{zho97} \\
 & OD-La214  & $S_{ab}$ & \onlinecite{yu96} \\
 & CeD-NdSr214 & $S$ & \onlinecite{amb02} \\
 & SrD-LaNd214 & $S$ & \onlinecite{tak00} \\
 & CD-Bi2201 & $S$ & \onlinecite{smi92,dev90} \\
\hline
 2  & CaD-Y1236 & $S$ & \onlinecite{hon04} \\
 & OD-Hg1212 & $S$ & \onlinecite{fuk97} \\
 & CD-HgFe1212 & $S$ & \onlinecite{kan05} \\
\hline 
 3 & OD-Hg1223 & $S$ & \onlinecite{fuk97} \\  
   & (underdoped) &    &   \\  
\end{tabular}
\end{ruledtabular}
\end{table}

The pseudogap was first noticed as the temperature showing a broad maximum in ($T_1T$)$^{-1}$ vs $T$ curve.\cite{tak91} The characteristic temperatures are observed as a broad maximum in the temperature dependence of the $S$ vs $T$,\cite{ber96} and $\gamma$ vs $T$.\cite{lor93} $S$($T$) can be scaled by $S$($T_S^*$) and $T_S^*$.\cite{hon04} The resistive pseudogap temperature ($T_{\rho}^*$) is defined as a temperature where the resistivity bends downward from the linear temperature dependence at the high temperature.\cite{ito93} The similar characteristic temperatures are observed also in $\chi$ vs $T$.\cite{oda97} The pseudogap by the QPR is observed as the gap-like behavior in substantial transient change of the optical transmission or reflection induced by ultrashort laser pulse photoexcitation.\cite{kab99} The ARPES and tunneling experiments provide us with the characteristic energies and temperatures, such as the peak and hump energies observed in the energy-dispersion at a fixed temperature and the temperature dependence of the energy-dispersion curve, respectively.\cite{dam03} The ERS give as the coherent and two-magnon peaks.\cite{sug03} In the NMR knight shift, $T_{mK}^*$ is a temperature where the constant $K_c$ at high temperature bends downward, and $T_K^*$ is a temperature where the linear $K_c$ below $T_{mK}^*$ bends downward.\cite{ish98,ito98} Recently the resistivity curvature mapping based on the data of in-plane resistivity up to 300 K showed that there are two inflection points, the upper inflection point and the lower inflection point, which are identified in the $\rho$ vs $T$ curve far above $T_c$.\cite{and04} Therefore, there are various characteristic temperatures and/or energies reported in the literature. Our goal is to see if we can put all of them into one unified phase diagram.

\section{Results and Discussion}

\subsection{Universal hole-doping scale}

\begin{figure}[t]
\includegraphics[scale=0.5]{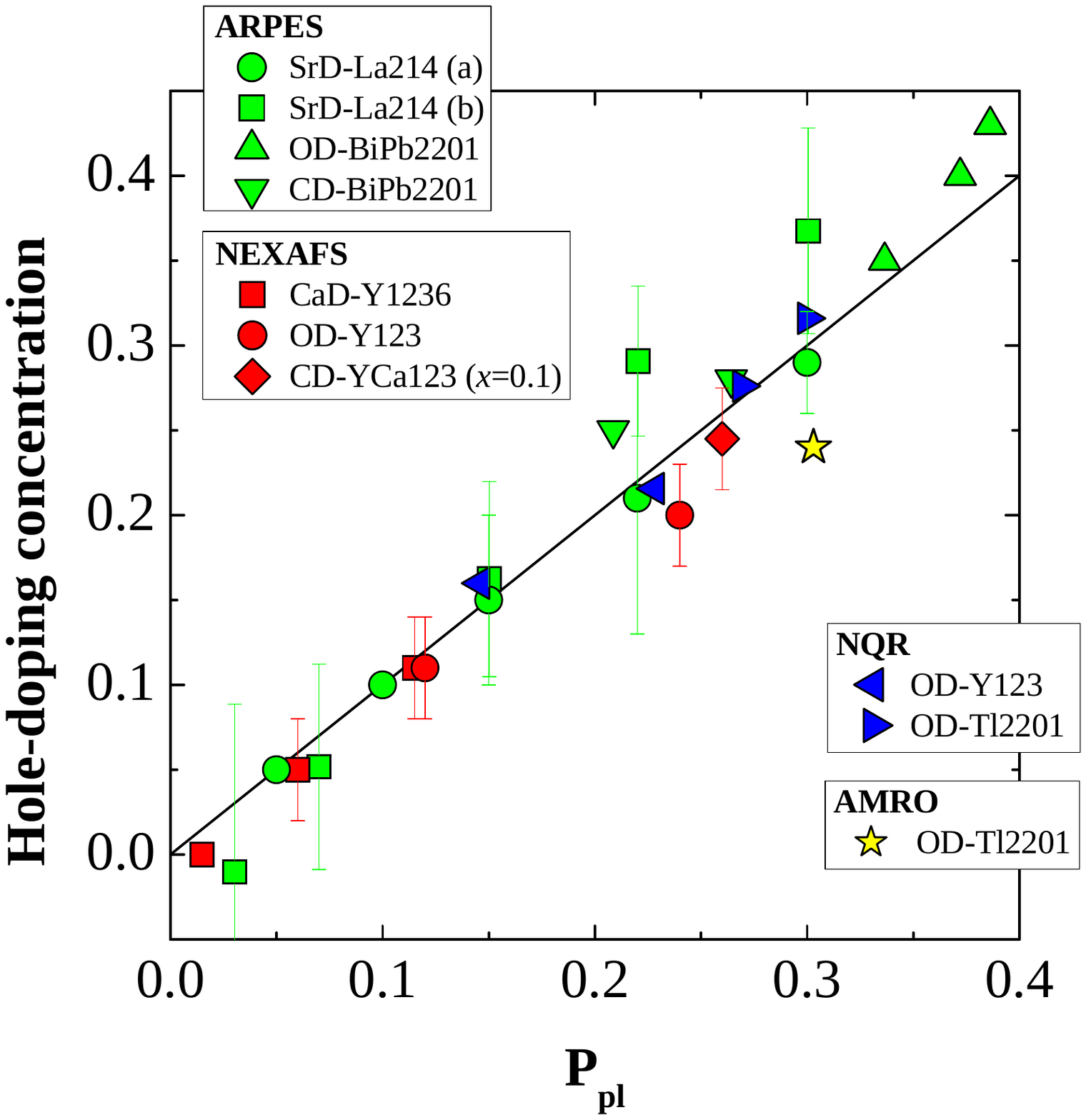}
\caption{\label{fig3} (Color online) Hole-doping concentration determined by various techniques as a function of $P_{pl}$. The plotted data are are summarized in Table~\ref{tab:table5}.}
\end{figure}

First of all, we demonstrate how the hole-doping scale based on the $S^{290}$ is effective and universal. In Figs.\ \ref{fig2}(a) and (b), we plot $S^{290}$ of sintered sample and $S_{ab}^{290}$ of the single crystal as a function of $P_{pl}$, together with previously reported data.\cite{hon04} $S^{290}$ ($\geq$ 7 $\mu$V/K) on the upper panel is plotted on a logarithmic scale, while $S^{290}$ ($<$ 7 $\mu$V/K) on the lower panel is plotted on a linear scale. In Figs.\ \ref{fig2}(a) and (b), the five single-layer, one double-layer, and one triple-layer HTSs are the newly added data points. They have been plotted with the previous reported SrD-La214 and CaD-Y1236. The plotted data are listed in Table~\ref{tab:table4}. $P_{pl}$ of SrD-La214 without excess oxygen is equal to Sr-content.\cite{tor88} The $P_{pl}$ of CaD-Y1236, which the oxygen-content was determined to be 6 by the iodometric titration in Ar gas,\cite{hon04} can be ambiguously and directly determined as a half of Ca-content, since the CaD-Y1236 has the isolated Cu layer in stead of CuO chain. In fact, it is shown by the O $1s$ and Cu $2p$ NEXAFS experiment that the holes introduced by replacing Y$^{3+}$ with Ca$^{2+}$ appear solely in the CuO$_2$ planes without affecting the isolated Cu layers in the CaD-Y1236.\cite{mer98} $P_{pl}$ for the other materials were determined from the copper valency measured by the iodometric titration for the OD-La214 \cite{yu96} and CD-Bi2201,\cite{dev90,smi92} and the double iodometric titration for the OD-Hg1212,\cite{fuk97} CD-HgFe1212, \cite{kan05} and OD-Hg1223.\cite{fuk97} The error of $P_{pl}$ is mainly coming from the oxygen-deficient ($\delta$). For the double- and triple-layer HTSs, the error of $\delta$ was below 0.01.\cite{hon04,fuk97,kan05} For the SrD-La214, the oxygen deficient is estimated to be $\sim$0.005, according to the result of Radaelli $et$ $al$.\cite{rad94} These error of $P_{pl}$ can be estimated to be below 0.01. For the CD-Bi2201, the error of $\delta$ is $\sim$0.02,\cite{dev90,smi92} and therefore the error of $P_{pl}$ is $\sim$0.04. Noticed that the plotted data follow the universal $S^{290}$($P_{pl}$)-curve proposed in Ref.\ \onlinecite{hon04}, which is irrespective of the nature of dopant, $n_{layer}$, the structure and the sample quality, namely, single crystal or not. It is also independent from whether the CuO$_2$ plane is surrounded by the octahedral or pyramidal oxygen coordination. For the SrD-La214, there is the upward deviation from the universal line at $P_{pl}$ $>$ 0.25. This deviation is considered to be due to the oxygen-deficient that was reported to be significant over $x$ = $P_{pl}$ $\sim$ 0.25.\cite{rad94} In the CeD-NdSr214, the upward deviation over $P_{pl}$ $\sim$ 0.15 from the universal line can be explained by the oxygen deficiency generating the hole deficient of $\sim$0.05 as pointed out in Ref.\ \onlinecite{amb02}. Accordingly, all plotted data lie in a shaded area around our universal $S^{290}$($P_{pl}$)-curve with the $P_{pl}$-accuracy of $\pm$0.01 within the reported error. Therefore, the proposed universal $S^{290}$($P_{pl}$)-curve that purely based on the experimental grounds works well as the empirical intrinsic hole-scale for the HTS in the range of 0.01 $<$ $P_{pl}$ $<$ 0.34. In Fig.\ \ref{fig2}(c), we compare $P_{pl}$ with $P_{T_c}$. The solid line shows $P_{pl}$ as a function of $P_{T_c}$. The broken line shows $P_{pl}$ = $P_{T_c}$. The quantitative difference between the $P_{pl}$ scale and $P_{T_c}$ scale becomes clear in Fig.\ \ref{fig2}(c). In addition we used this relation for the conversion from $P_{T_c}$ into $P_{pl}$ when the data plotted here have the $P_{T_c}$ without TEP.

Next, we compare our universal scale based on the $S^{290}$ to that determined by other techniques. The hole-doping concentration by ARPES ($P_{ARPES}$)  is deduced from the area of the experimental Fermi surface (FS). The planar hole-doping concentration is also determined by NEXAFS ($P_{NEXAFS}$), by NQR ($P_{NQR}$) and by AMRO ($P_{AMRO}$). In Fig.\ \ref{fig3}, we plot the $P_{ARPES}$, $P_{NEXAFS}$, $P_{NQR}$ and $P_{AMRO}$ as a function of $P_{pl}$. The plotted data are summarized in Table~\ref{tab:table5}. It can be clearly seen that the $P_{pl}$ determined by TEP is quite consistent with $P_{NEXAFS}$ and $P_{NQR}$. Although there is a slight scattering, $P_{pl}$ is also consistent with $P_{ARPES}$ and $P_{AMRO}$. Thus, our $P_{pl}$-scale is consistent with above other scales. Accordingly, the present $p_u$ scale is also intrinsically consistent with the hole-doping concentrations determined by the above techniques.

\begin{table}[t]
\caption{\label{tab:table5}The data plotted in Fig.\ \ref{fig3}.}
\begin{ruledtabular}
\begin{tabular}{llcl}
 \quad Probe & \quad HTS & $P_{pl}$ & Ref. \qquad \\ \hline
 \quad ARPES & SrD-La214 (a) &  & \onlinecite{ino02} \\
       & SrD-La214 (b) &  & \onlinecite{yos06} \\
       & OD-BiPb2201 &  & \onlinecite{kon05} \\
     & CD-BiPb2201 &  & \onlinecite{kon05}
\\
\hline
\quad NEXAFS & CaD-Y1236 &  & \onlinecite{mer98} \\
       & OD-Y123 & I & \onlinecite{mer98} \\
       & CD-YCa123 ($x$=0.1) & I & \onlinecite{mer98} \\
\hline
\quad NQR    & OD-Y123 & I & \onlinecite{kot01} \\
       & OD-Tl2201 & I & \onlinecite{kot01} \\
\hline 
\quad AMRO & OD-Tl2201 & I &\onlinecite{hus03} \\ 
\end{tabular}
\end{ruledtabular}
\end{table}

\begin{figure}[b]
\includegraphics[scale=0.5]{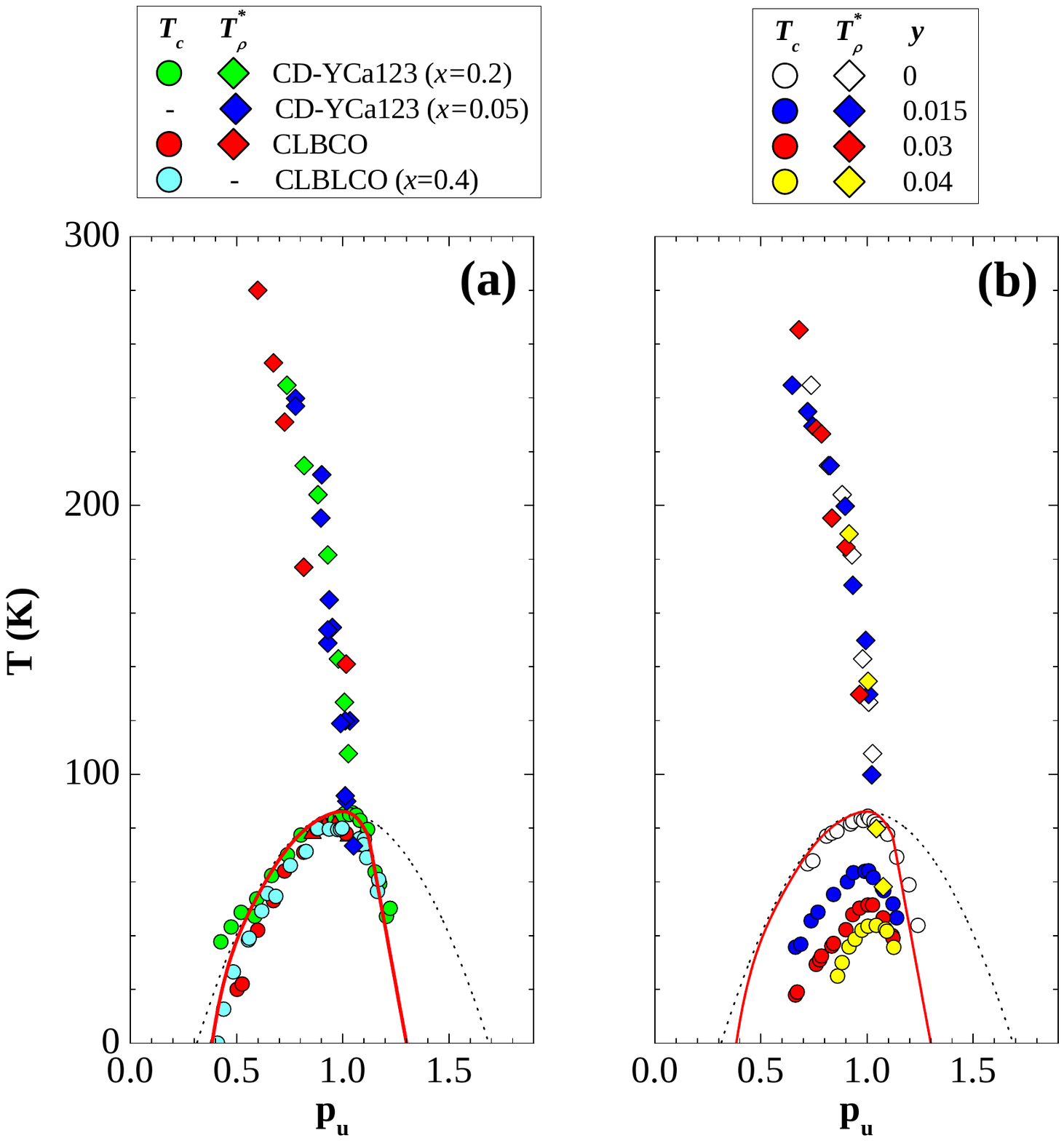}
\caption{\label{fig4} (Color online) $T_c$ and $T_{\rho}^*$ as a function of $p_u$ for (a) the OD-Y123-related materials and (b) Y$_{0.8}$Ca$_{0.2}$Ba$_2$(Cu$_{1-y}$Zn$_y$)$_3$O$_{6+\delta}$. The plotted data are summarized in Table~\ref{tab:table6}. The solid line is a half-dome-shaped $T_c$-curve with $T_c^{max}$ = 86 K. The dotted line comes from the equation (\ref{form1}) with $T_c^{max}$ = 86 K.}
\end{figure}

\begin{table}[b]
\caption{\label{tab:table6}The data plotted in Figs.\ \ref{fig4}(a) and (b).}
\begin{ruledtabular}
\begin{tabular}{llcl}
 \quad Fig. & \quad HTS & $P_{pl}$ & Ref(s). \quad \\ \hline
 \quad 4(a)  & CD-YCa123 ($x$=0.2) &  & \onlinecite{coo00,ber96} \\
       & CD-YCa123 ($x$=0.05) & II & \onlinecite{naq05} \\
       & CLBLCO ($x$=0.4) & & \onlinecite{kni99} \\
       & CLBCO & & \onlinecite{hay96} \\
\hline
 \quad 4(b)  & Y$_{0.8}$Ca$_{0.2}$Ba$_2$(Cu$_{1-y}$Zn$_y$)$_3$O$_{6+\delta}$ & II & \onlinecite{coo00} \\
       & (0 $\leq$ $y$ $\leq$ 0.04) & & \\
\end{tabular}
\end{ruledtabular}
\end{table}

\subsection{Asymmetric half-dome-shaped $T_c$ curve}

In Fig.\ \ref{fig4}(a), we plot the $T_c$ and $T_{\rho}^*$ of all the OD-Y123 related materials which do not have significant contribution of CuO chain as a function of $p_u$. It includes CD-YCa123,\cite{naq05,coo00,ber96} CLBLCO,\cite{kni99} and CLBCO.\cite{hay96} First, we note that the $T_c$($p_u$) does not follow the well-known superconducting dome, as shown as a dotted line in Fig.\ \ref{fig4}(a), instead, it follows an asymmetric half-dome-shaped curve shown as a solid line. Although $T_c$ in the underdoped regime basically follows the superconducting dome, $T_c$ in the overdoped regime decreases much more rapidly. In Fig.\ \ref{fig4}(a), $T_{\rho}^*$ decreases with doping, smoothly merges into the half-dome-shaped $T_c$ curve and finally tends to reach an end point located at ($p_u$,$T$) = (1.3,0). Therefore, in contrast to the proposal that the $T_{\rho}^*$ curve crosses the $T_c$ curve,\cite{tal01} the $T_{\rho}^*$ curve smoothly merges into the $T_c$ curve in the overdoped regime. In Fig.\ \ref{fig4}(b), we plot $T_c$ and $T_{\rho}^*$ as a function of $p_u$ for Y$_{0.8}$Ca$_{0.2}$Ba$_2$(Cu$_{1-y}$Zn$_y$)$_3$O$_{6+\delta}$ for 0 $\leq$ $y$ $\leq$ 0.04.\cite{naq05}, Although  $T_{\rho}^*$($p_u$) slightly depends on Zn-content in the overdoped regime, $T_{\rho}^*$($p_u$) again tends to merge into $T_c$($p_u$) at the overdoped regime. This should be compared to the original plot, based on the $P_{T_c}$-scale, in which $T_{\rho}^*$ crossed the superconducting dome and reached zero at a proposed QCP ($P_{Tc}$ = 0.19).\cite{naq05} Accordingly, the crossing was a artifact that came from two sources; one is the use of a hole-scale that failed to taking into account the differences in dimensionality of different physical properties, namely, the two-dimensional $T_{\rho}^*$ vs\ three-dimensional \textit{\textbf{$T_c$}}, {\bf and} the other is that the $T_c$ curve for the majority of HTS follows the asymmetric dome-shaped curve, \textit{\textbf{only}} SrD-La214 follows the symmetric dome-shaped $T_c$curve or superconducting dome.

\begin{figure}[b]
\includegraphics[scale=0.5]{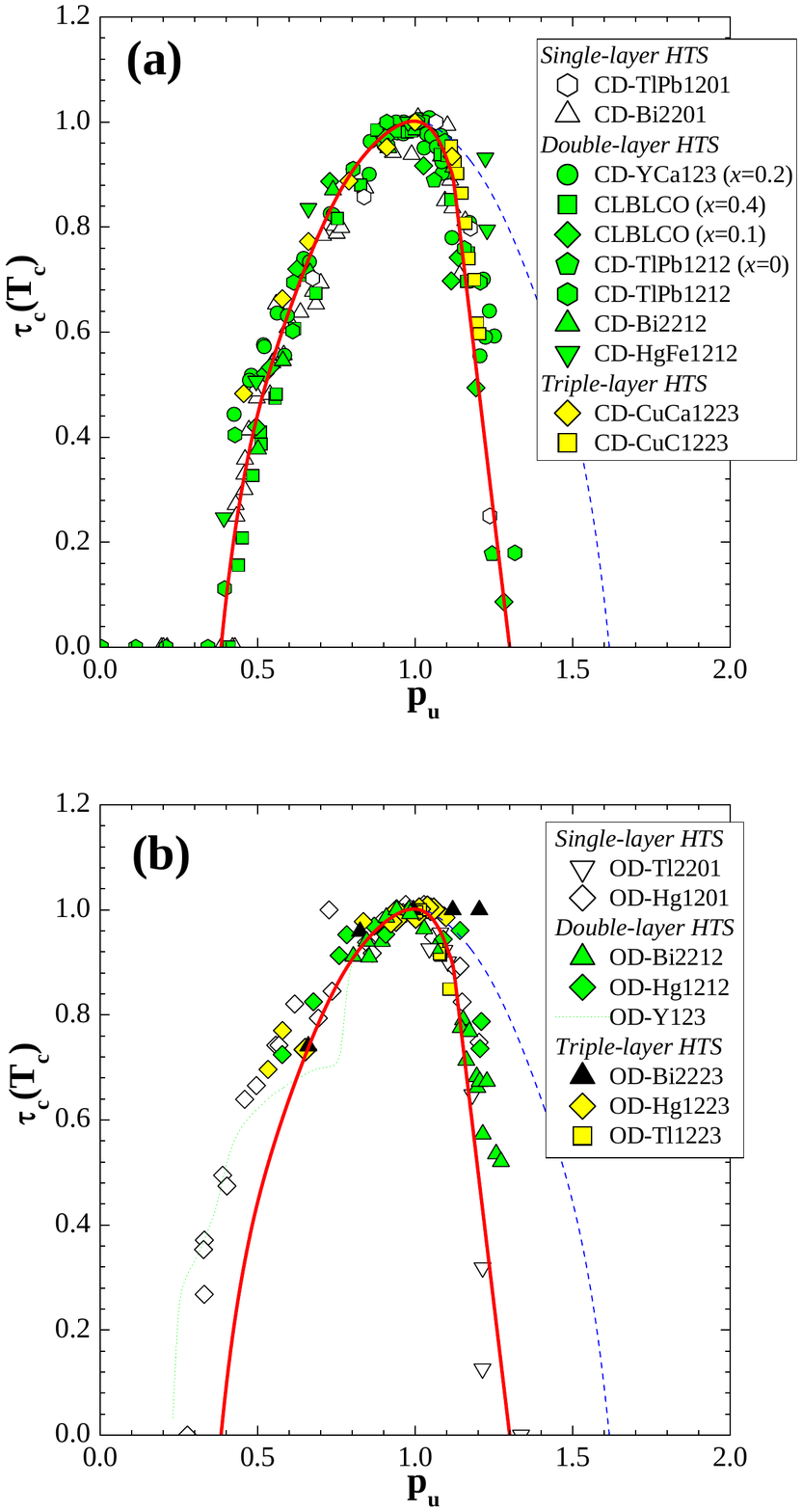}
\caption{\label{fig5} (Color online) Extended unified electronic phase diagram plotted as $\tau_c$($T_c$) vs $p_u$ for (a) the cation and oxygen co-doped HTS's and (b) the purely oxygen doped HTS. The plotted data are summarized in Table~\ref{tab:table7}. The solid and broken lines are an asymmetric half-dome-shaped $T_c$ curve and our superconducting dome, respectively. The dotted line is the $T_c$ curve for OD-Y123.\cite{hon07}}
\end{figure}

$\tau_c$($T_c$) vs $p_u$ plot for the cation and oxygen co-doped HTS and the purely oxygen-doped HTS are shown in Figs.\ \ref{fig5}(a) and (b), respectively. For comparison, $\tau_c$($T_c$) vs $P_{pl}$ curve of OD-Y123 reported in Ref.\ \onlinecite{hon07} is also plotted in Fig.\ \ref{fig5}(b). $T_c^{max.}$ and $P_{pl}^{opt.}$ are summarized in Table~\ref{tab:table7}. The CD-HTSs follow the present asymmetric half-dome-shaped $T_c$-curve. $\tau_c$($T_c$) vs $p_u$ curve of the single-layer OD-Tl2201, which behaves differently from that of the other in the plot of $T_c$ vs $P_{3D}$ as shown in Fig.\ \ref{fig1}(b),\cite{hon06} actually follows the asymmetric half-dome shaped $T_c$-curve. The other overdoped OD-HTSs also follow the half-dome-shaped $T_c$ curve. Note that $T_c$ of the underdoped OD-HTS is slightly enhanced from the half-dome shaped $T_c$-curve and $T_c$ appears at a lower $p_u$. The OD-Y123 also shows the similar trend, although it is influenced by the CuO chain ordering.\cite{hon07} We attribute this to the influence of the soft oxygen dopants.\cite{lor02} Thus, opposite to the common belief, the $\tau_c$($T_c$) vs $p_u$ phase diagram of the majority of HTSs follow the asymmetric half-dome shaped $T_c$ curve. Noticed that the asymmetric half-dome-shaped $T_c$ curve goes to zero at $p_u$ $\sim$ 1.3. It is interesting to point out that if we take $P_{pl}^{opt.}$ to be universally equal to 0.16, as assumed in the $P_{T_c}$ scale, then the critical $p_u$ $\sim$ 1.3 corresponds to $P_{T_c}$ $\sim$ 0.2 in the $P_{T_c}$ scale. This value is very close to the proposed QCP ($P_{T_c}$ = 0.19) identified by various experiments on the $P_{T_c}$-scale.\cite{tal01,naq05} Therefore this critical doping concentration is not located inside the superconducting phase and, physically, it is the doping concentration where all the phenomenology of high $T_c$ ceases to exist and the ground state becomes a conventional Fermi liquid (FL) for $p_u$ $>$ 1.3.

\begin{table}
\caption{\label{tab:table7}The $T_c^{max.}$ and $P_{pl}^{opt.}$ for the HTSs plotted in Figs.\ \ref{fig5}(a) and\ \ref{fig5}(b).}
\begin{ruledtabular}
\begin{tabular}{llclcl}
 Fig. & \qquad HTS & $T_c^{max.}$(K) & $P_{pl}^{opt.}$ & $P_{pl}$ & Ref(s). \\ \hline
 5(a) & (Single-layer HTS) \\
 & CD-TlPb1201 & 50 & 0.25 & & \onlinecite{sub94} \\
 & CD-Bi2201 & 35.5 & 0.28 & &  \onlinecite{and00} \\
 & CD-Bi2201 & 33 & 0.28 & &  \onlinecite{oka06} \\
\cline{2-6}
 & (Double-layer HTS) \\
 & CD-YCa123 ($x$=0.2) & 85 & 0.237 & &  \onlinecite{coo00,lor98} \\
 & CD-YCa123 ($x$=0.2) & 81 & 0.238 & II &  \onlinecite{naq05} \\
 & CD-YCa123 ($x$=0.2) & 85.5 & 0.238 & &  \onlinecite{ber96} \\
 & CLBLCO ($x$=0.4) & 81 & 0.235 & &  \onlinecite{kni99} \\
 & CLBLCO ($x$=0.1) & 57.7 & 0.205 & &  \onlinecite{kni99} \\ 
 & CD-TlPb1212 ($x$=0) & - $^{\text{a}}$ & 0.235 & &  \onlinecite{ber96} \\
 & CD-TlPb1212 & 94 & 0.235 & &  \onlinecite{sub92} \\
 & CD-Tl1212 & 90 & 0.235 & &  \onlinecite{mar95} \\
 & CD-Bi2212 & 92 & 0.236 & &  \onlinecite{ako98} \\
 & CD-Bi2212 & 81 & 0.238 & &  \onlinecite{man96} \\
 & CD-HgFe1212 & 73 & 0.227 & &  \onlinecite{kan05} \\
\cline{2-6}
 & (Triple-layer HTS) \\
 & CD-CuCa1223 & 122 & 0.248 & &  \onlinecite{cao97} \\
 & CD-CuC1223 & 110 & 0.215 & &  \onlinecite{cao97} \\
 \hline
5(b) & (Single-layer HTS) \\
 & OD-Tl2201 & 93$^{\text{b}}$ & 0.25$^{\text{b}}$ & &  \onlinecite{obe92,shu93,mar95} \\
 & OD-Hg1201 & 97 & 0.235 & &  \onlinecite{yam00,wil00} \\
\cline{2-6}
 & (Double-layer HTS) \\
 & OD-Bi2212 & 92 & 0.238 & &  \onlinecite{obe92} \\
 & OD-Hg1212 & 127 & 0.227 & &  \onlinecite{fuk97} \\
 & OD-Hg1212 & 125 & 0.227 & &  \onlinecite{coh99} \\
\cline{2-6}
 & (Triple-layer HTS) \\
 & OD-Bi2223 & 108 & 0.215 & &  \onlinecite{fuj02} \\
 & OD-Hg1223 & 135 & 0.215 & &  \onlinecite{fuk97} \\
 & OD-Hg1223 & 138 & 0.215 & &  \onlinecite{sub95} \\
 & OD-Tl1223 & 128 & 0.23 & &  \onlinecite{mik06} \\

\end{tabular}
\end{ruledtabular}
\footnotetext[1]{The $T_c$/$T_c^{max.}$ was reported.}
\footnotetext[2]{We use the reported highest $T_c$ = 93 K as $T_c^{max.}$.\cite{wag97} From the plot of $T_c$ vs $P_{pl}$ in Fig.\ \ref{fig1}(a), the optimal $P_{pl}$ is estimated to be $\sim$0.25. The detail is in the text.}
\end{table}

In the overdoped triple-layer HTS, the charge density of the inner and outer planes were reported to be inhomogeneous.\cite{tok99} This is consistent with the $\tau_c$($T_c$) vs $p_u$ behavior of OD-Bi2223, black triangles in Fig.\ \ref{fig5}(b), that $T_c$ shows a flat region in the overdoped regime.\cite{fuj02} However, the $\tau_c$($T_c$) vs $p_u$ behaviors of CD-CuCa1223,\cite{cao97} and CD-CuC1223 \cite{cao97} plotted into Fig.\ \ref{fig5}(a) show the same trend as that of the single- and double-layer CD-HTSs. $\tau_c$($T_c$) vs $p_u$ of OD-Tl1223,\cite{mik06} and OD-Hg1223 \cite{fuk97,sub95} plotted in Fig.\ \ref{fig5}(b) also show the identical trend as that of the single- and double-layer OD-HTSs. Accordingly, although counter-intuitive, the charge density of the inner and outer planes of these materials is expected to be the same.

\subsection{Unified electroni phase diagram}

\begin{figure*}
\includegraphics[scale=0.6]{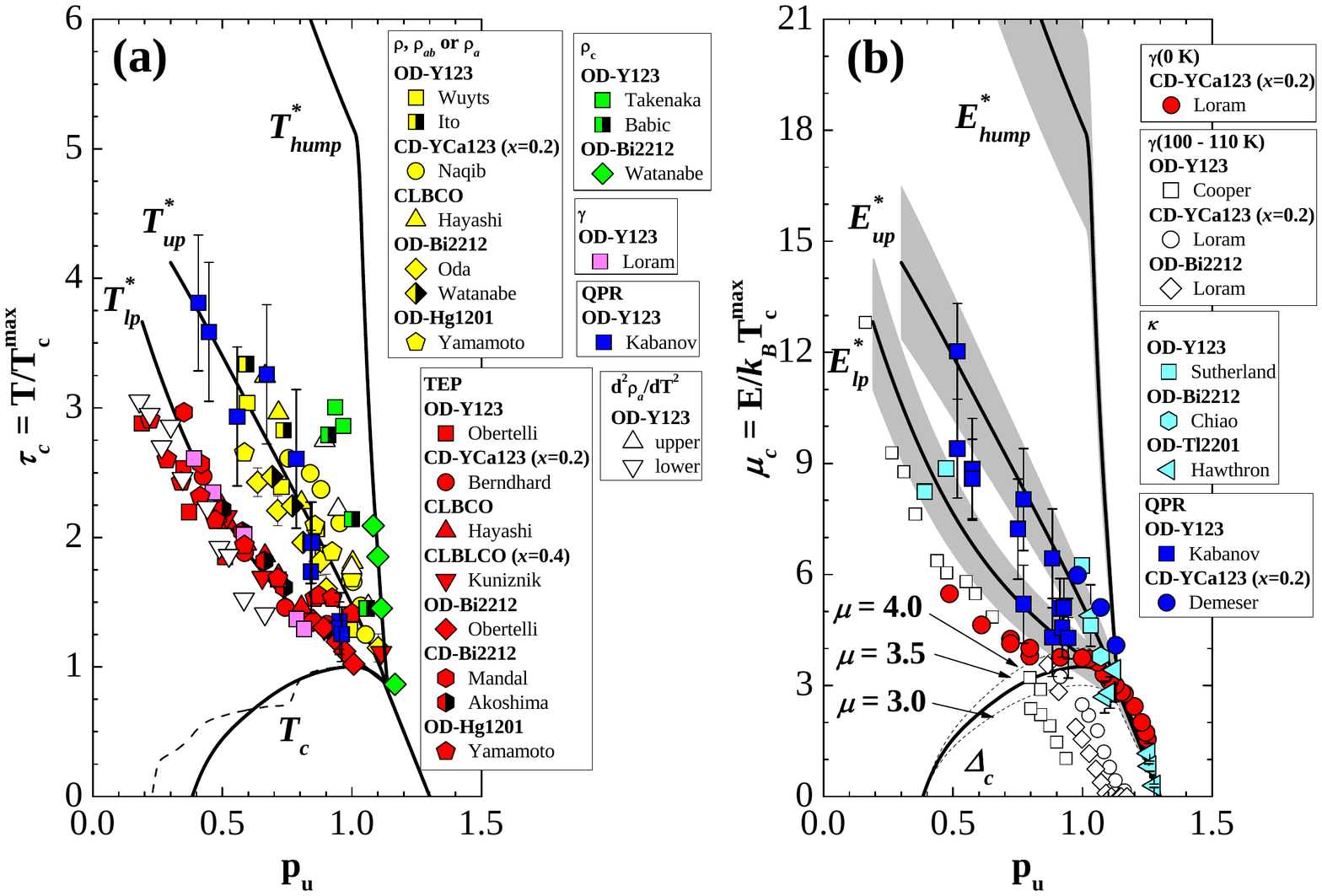}

\caption{\label{fig6} (Color online) Unified electronic phase diagram for the single- and double-layer HTS's with $T_c^{max}$ $\sim$ 90 K. The temperature- and energy-scale for the pseudogap and superconducting gap obtained from transport properties are summarized in Figs.\ \ref{fig6}(a) and\ \ref{fig6}(b), from the spectroscopy properties in Figs.\ \ref{fig6}(c) and (d), and from NMR, QPR and scattering properties in Figs.\ \ref{fig6}(e) and\ \ref{fig6}(f), respectively. For Fig.\ \ref{fig6}(a) and\ \ref{fig6} (b), the plotted data are summarized in the Table~\ref{tab:table8}. For Figs.\ \ref{fig6}(c) -\ \ref{fig6}(f), the plotted data are summarized in the Table~\ref{tab:table9}. The $E_{hump}^*$, $T_{up}^*$, $T_{lp}^*$, and $T_c$ curves are directly determined from the plotted data. The $T_{hump}^*$, $E_{up}^*$, $E_{lp}^*$, and $\Delta_c$ curves are calculated from the $E_{hump}^*$, $T_{up}^*$, $T_{lp}^*$, and $T_c$-curves using a relation of $T$ = $E$/$zk_B$ or $E$ = $zk_BT$, respectively. The broken lines show the $T_c$ curve or $\Delta_c$ curve the for OD-Y123. In the energy-scale, the solid lines corresponds to $z$ = 3.5 and gray zone shows the energy range from 3$k_BT$ to 4$k_BT$. }
\end{figure*}

\begin{figure*}
\includegraphics[scale=0.6]{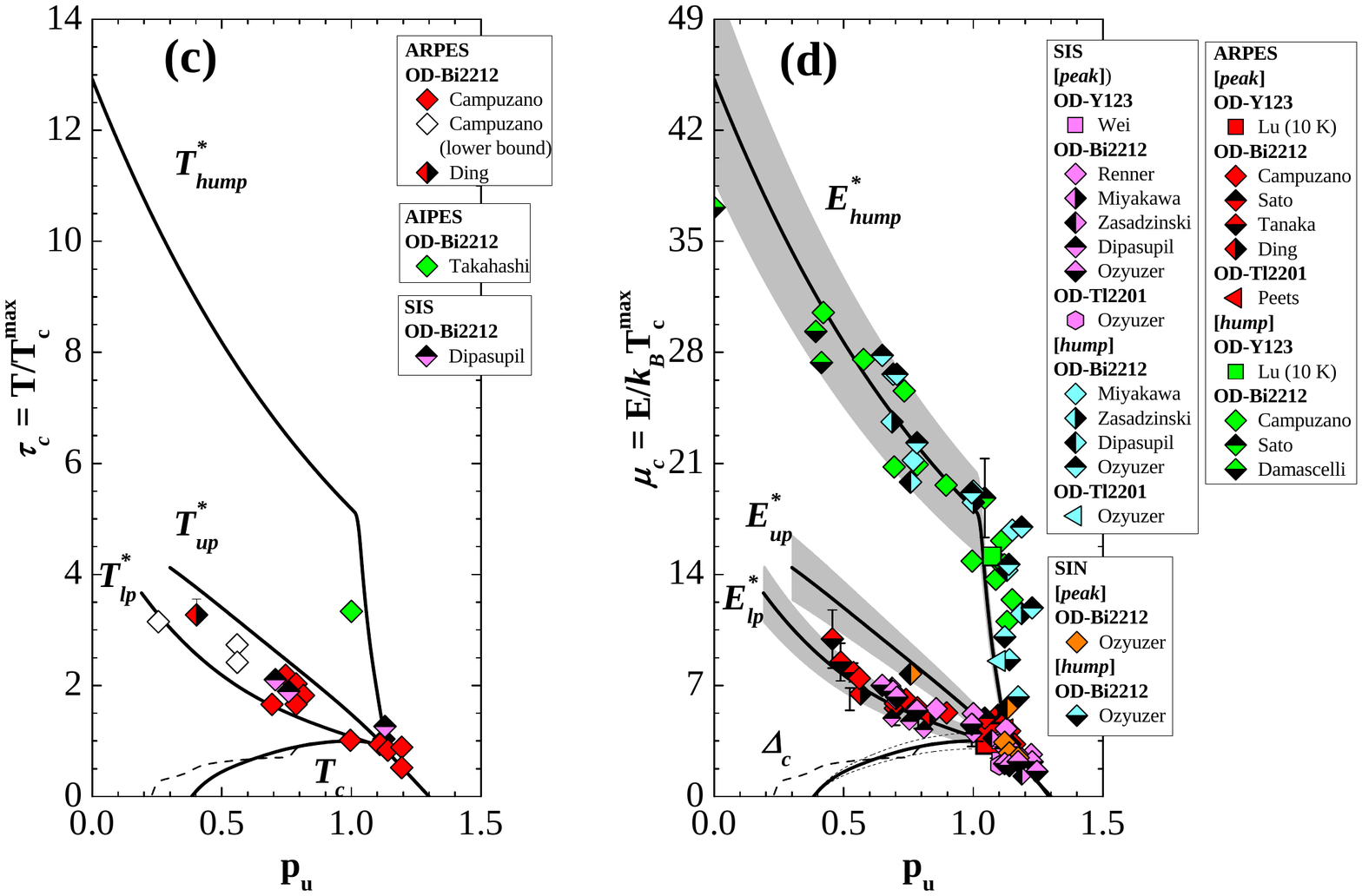}
\includegraphics[scale=0.6]{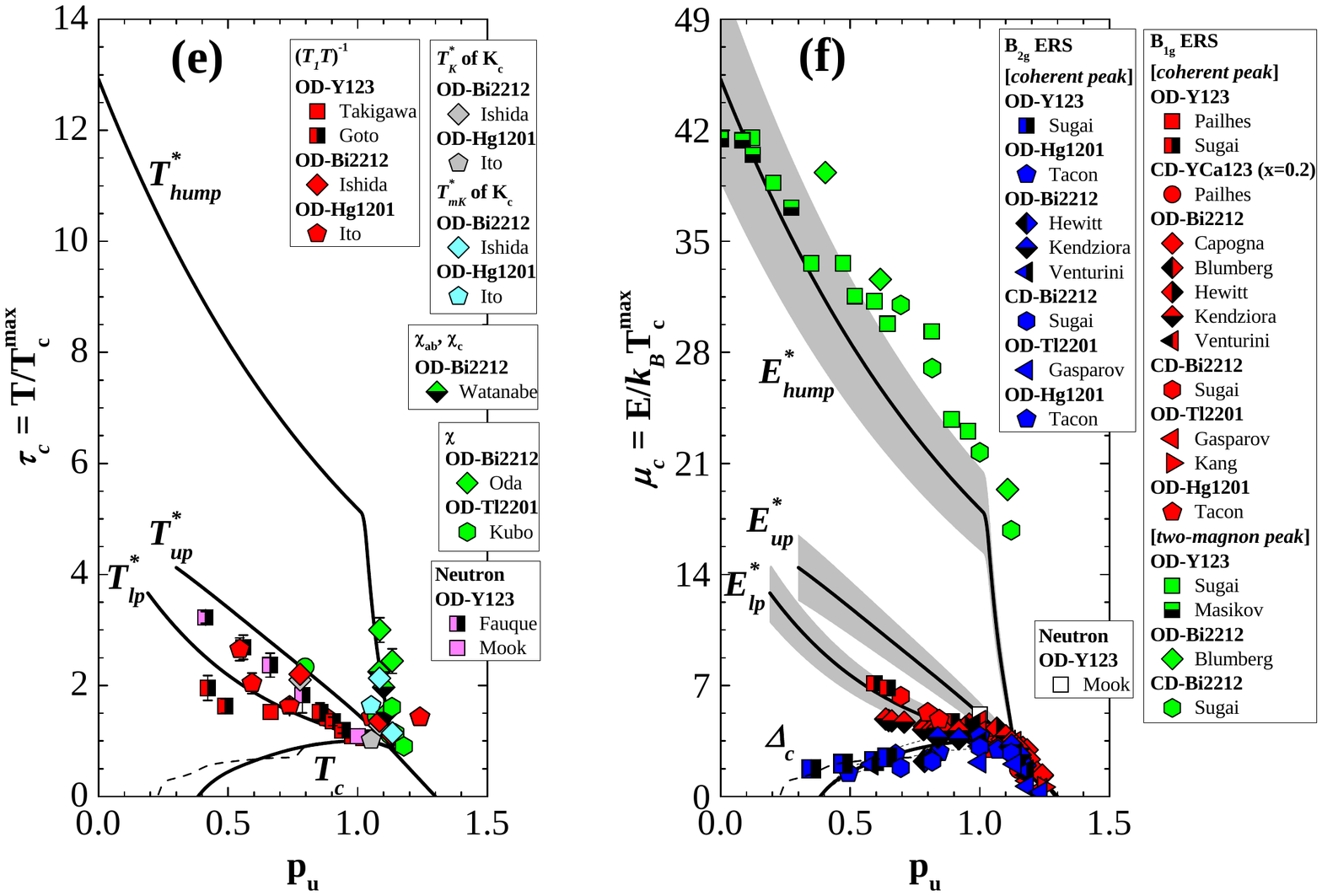}
\end{figure*}

\begin{table}
\caption{\label{tab:table8}The HTSs plotted in Figs.\ \ref{fig6}(a) and\ \ref{fig6}(b).}
\begin{ruledtabular}
\begin{tabular}{lllcl}
 Fig. & Probe & HTS & $P_{pl}$ & Ref(s). \\ \hline
 6(a) & $\rho$      & OD-Y123           & I & \onlinecite{wuy96} \\
      &             & CD-YCa123 ($x$=0.2) & II & \onlinecite{naq05} \\
      &             & CLBCO            &  & \onlinecite{hay96} \\
      &             & OD-Hg1201         &     & \onlinecite{yam00} \\
\cline{2-5}
      & $\rho_a$    & CD-Bi2212         &     & \onlinecite{wat00} \\
\cline{2-5}
      & $\rho_{ab}$ & OD-Y123           & I & \onlinecite{ito93} \\
      &             & OD-Bi2212         & I & \onlinecite{oda97} \\
\cline{2-5}
      & $\rho_c$    & OD-Y123           & I & \onlinecite{tak94,bav99} \\
      &             & OD-Bi2212         & I & \onlinecite{wat00} \\
\cline{2-5}
      & $d^2\rho_a/dT^2$ & OD-Y123         & I & \onlinecite{and04} \\
\cline{2-5}
      & TEP         & OD-Y123           &     & \onlinecite{obe92} \\
      &             & CD-YCa123 ($x$=0.2) &     & \onlinecite{ber96} \\
      &             & CLBCO             &     & \onlinecite{hay96} \\
      &             & CLBLCO ($x$=0.4)  &     & \onlinecite{kni99} \\
      &             & OD-Bi2212         &     & \onlinecite{obe92} \\
      &             & CD-Bi2212         &     & \onlinecite{man96,ako98} \\
      &             & OD-Hg1201         &     & \onlinecite{yam00}\\
\cline{2-5}
      & $\gamma$    & OD-Y123         & I & \onlinecite{lor93} \\
\cline{2-5}
      & QPR         & OD-Y123           & I & \onlinecite{kab99} \\
 \hline
6(b)  & $\gamma$    & CD-YCa123 ($x$=0.2) & I & \onlinecite{low97} \\
      &             & OD-Y123           &     & \onlinecite{coo96} \\
      &             & OD-Bi2212         & I & \onlinecite{low00} \\
\cline{2-5}
      & $\kappa$     & OD-Y123           & I & \onlinecite{sut03} \\
      &              & OD-Bi2212         & I & \onlinecite{cha00} \\
      &             & OD-Tl2201         & I & \onlinecite{haw07} \\
\cline{2-5}
      & QPR         & OD-Y123           & I & \onlinecite{kab99} \\
      &             & CD-YCa123 ($x$=0.2) & I & \onlinecite{dem99} \\
\end{tabular}
\end{ruledtabular}
\end{table}

\begin{table}
\caption{\label{tab:table9}The HTSs plotted in Figs.\ \ref{fig6}(c) -\ \ref{fig6}(f).}
\begin{ruledtabular}
\begin{tabular}{lllcl}
 Fig. & Probe & HTS & $P_{pl}$ & Ref(s). \\ \hline
6(c)  & ARPES       & OD-Bi2212         & I & \onlinecite{cam99,din96,sat02} \\
\cline{2-5}
      & AIPES       & OD-Bi2212         & I & \onlinecite{tak01} \\
\cline{2-5}
      & SIS         & OD-Bi2212         & I & \onlinecite{dip02} \\
 \hline
6(d)  & ARPES       & OD-Y123           & I & \onlinecite{lu01} \\
      &             & OD-Bi2212         & I & \onlinecite{cam99,sat02,tan06,din01,dam03} \\
      &             & OD-Tl2201         & I & \onlinecite{pee07} \\
\cline{2-5}
      & SIS         & OD-Y123           & I & \onlinecite{wei98} \\
      &             & OD-Bi2212         & I & \onlinecite{ren98,miy98,zas01,dip02,ozy00} \\
      &             & OD-Tl2201         & I & \onlinecite{ozy99} \\
\cline{2-5}
      & SIN         & OD-Bi2212         & I & \onlinecite{ozy00} \\
 \hline
6(e)  & ($T_1T$)$^{-1}$  & OD-Y123      & I & \onlinecite{tak91,got96} \\
      &             & OD-Bi2212         & I & \onlinecite{ish98} \\
      &             & OD-Hg1201         & I & \onlinecite{ito98} \\
\cline{2-5}
      & $K_c$       & OD-Bi2212 & I & \onlinecite{ish98} \\
      &             & OD-Hg1201         & I & \onlinecite{ito98} \\
\cline{2-5}
      & neutron     & OD-Y123           & I & \onlinecite{faq06,moo93} \\
\cline{2-5}
      & $\chi$      & OD-Bi2212         & I & \onlinecite{oda97} \\
      &             & OD-Tl2201         & I & \onlinecite{kub91} \\
\cline{2-5}
      & $\chi_{ab}$ and $\chi_c$ & OD-Bi2212  & I & \onlinecite{wat00} \\
 \hline
6(f)  & $B_{1g}$ ERS & OD-Y123          & I & \onlinecite{pai06,sug03,mak94} \\
      &             & CD-YCa123 ($x$=0.2) & I & \onlinecite{pai06} \\
      &             & OD-Bi2212         & I & \onlinecite{cap07,blu97,hew02,ken95,ven02} \\
      &             & CD-Bi2212         & I & \onlinecite{sug00} \\
      &             & OD-Tl2201         & I & \onlinecite{kan96} \\
      &             & OD-Hg1201         & I & \onlinecite{tac06} \\
\cline{2-5}
      & $B_{2g}$ ERS & OD-Y123          & I & \onlinecite{sug03} \\
      &             & OD-Bi2212         & I & \onlinecite{hew02,ken95,ven02} \\
      &             & CD-Bi2212         & I & \onlinecite{sug03} \\
      &             & OD-Tl2201         & I & \onlinecite{gas98} \\
      &             & OD-Hg1201         & I & \onlinecite{tac06} \\
\cline{2-5}
      & neutron     & OD-Y123           & I & \onlinecite{moo93} \\
\end{tabular}
\end{ruledtabular}
\end{table}

We now examine various characteristic temperatures and energies for HTSs that fall into the asymmetric half-dome-shaped $T_c$ curve. $T_c$($P_{pl}$) depends on the $n_{layer}$, while the $T^*$($P_{pl}$) is independent of it.\cite{hon04} Therefore, we group the single- and double-layer HTSs with same $T_c^{max}$ $\sim$ 90 K together. We found in Ref.\ \onlinecite{hon04} that the various characteristic temperatures or pseudogap temperatures can be separated into two groups of the lower pseudogap temperature ($T_{lp}^*$) and upper pseudogap temperature ($T_{up}^*$). $T_c$ and major characteristic temperatures, which includes $T_{lp}^*$ and $T_{up}^*$, are plotted on the reduced temperature-scale as a function of $p_u$ in Figs.\ \ref{fig6}(a), (c) and (e). The characteristic energies are plotted on a reduced energy-scale $\mu_c$($E$) $\equiv$ $E$/$k_BT_c^{max}$, where $k_B$ is Boltzmann's constant, as a function of $p_u$ in Figs.\ \ref{fig6}(b), (d) and (f). We will call the four solid curves from the top to bottom the hump temperature ($T_{hump}^*$), $T_{up}^*$, $T_{lp}^*$ and $T_c$ curves in the temperature-scale, and the hump energy ($E_{hump}^*$), the upper pseudogap energy ($E_{up}^*$), the lower pseudogap energy ($E_{lp}^*$) and $\Delta_c$-curves in the energy-scale, respectively. The $E_{hump}^*$, $T_{up}^*$, $T_{lp}^*$ and $T_c$-curves are directly determined from the plotted data. The $T_{hump}^*$, $E_{up}^*$, $E_{lp}^*$ and $\Delta_c$ curves are converted from the $E_{hump}^*$, $T_{up}^*$, $T_{lp}^*$ and $T_c$ curves using a relation of $T$ = $E$/$zk_B$ or $E$ = $zk_BT$ for each characteristic energy or temperature. In the energy-scale, the solid curves correspond to $z$ = 3.5 and gray zone shows the energy range from 3$k_BT$ to 4$k_BT$.

First, we summarize the characteristic temperatures and energies derived from the transport and thermodynamic properties in Figs.\ \ref{fig6}(a) and (b). Here, the plotted data are summarized in Table~\ref{tab:table8}. $T^*$'s determined from $TEP$ and $\gamma$ lie on the $T_{lp}^*$-curves, while $T^*$'s determined from $\rho$ and QPR lie on the $T_{up}^*$ curves, as reported in Ref.\ \onlinecite{hon04}. Accordingly, the upper pseudogap is identified by the $\rho$ and QPR, and the lower pseudogap is identified by $TEP$ and $\gamma$ experiments. However, $T^*$ determined from the $\rho_c$ tends to be higher than $T_{up}^*$, although the doping range is restricted. This may suggest a third pseudogap as already pointed out in Ref.\ \onlinecite{hon04}. This suggestion is further supported by the similar behavior derived by other probes in the temperature- and energy-scales. We plot the upper and lower inflection points of $\rho_a$ of OD-Y123 into the Fig.\ \ref{fig6}(a).\cite{and04} The lower and upper inflection points seem to lie on the $T_{lp}^*$ and $T_{up}^*$ curves, respectively.

For the characteristic energies, we use the data reported in the $\gamma$,\cite{low97,coo96,low00} $\kappa$,\cite{sut03,haw07,cha00} and QPR experiment.\cite{kab99,dem99} $E^*$'s determined from the QPR lie on the $E_{up}^*$ urve. $E^*$'s determined from $\kappa$ show up on either $E_{lp}^*$ or $E_{up}^*$ curve. In the overdoped side, these $E^*$'s clearly merge into the $\Delta_c$-curve. This indicates that there is no QCP inside the superconducting phase. In Fig.\ \ref{fig6}(b), the normal state gap, $E_{sh}^*$(110 K) and zero temperature superconducting gap, $E_{sh}^*$(0 K) determined by the specific heat measurement are plotted as the open symbols and solid circles, respectively.\cite{low97,coo96,low00} The $E_{sh}^*$(110 K) at $p_u$ $<$ 0.85 and $E_{sh}^*$(0 K) follows the $E_{lp}^*$- or $zk_BT_{lp}^*$-curve. However, $E_{sh}^*$(110 K) at $p_u$ $>$ 0.85 deviates downward from the $E_{lp}^*$-curve, crosses the $\Delta_c$-curve and finally goes to zero inside the $\Delta_c$-curve. The temperature of $\sim$110 K ($\tau_c$(110 K) = 110/90 $\sim$ 1.2) corresponds to the $T_{lp}^*$ at $p_u$ $\sim$ 0.85. The influence of the lower pseudogap on the extraction of $E^*$'s is clearly seen. The plotted $E_{sh}^*$(110 K) is the same data set used to support the existence of the QCP \textit{\textbf{inside}} the superconducting phase on the $P_{T_c}$ scale.\cite{tal01} Accordingly, the existence of the QCP inside the superconducting phase is extrinsic to high $T_c$.

We summarize the characteristic temperatures and energies derived from the spectroscopic measurements in Figs.\ \ref{fig6}(c) and (d). The plotted data are summarized in Table~\ref{tab:table9}. $T^*$ determined in the AIPES lies on $T_{hump}^*$.\cite{tak01} $T^*$ determined from the ARPES,\cite{din96,cam99,sat02} and SIS \cite{dip02} cannot be grouped into either $T_{up}^*$ or $T_{lp}^*$ curve, since they lie between $T_{up}^*$ and $T_{lp}^*$-curves. The $E^*$ determined in the ARPES,\cite{cam99,sat02,din01,lu01,tan06,pee07,dam03} and tunneling,\cite{dip02,ren98,miy98,wei98,zas01,ozy99,ozy00} are plotted in Fig.\ \ref{fig6}(d): the peak and hump energies observed in ARPES and tunneling lie on the $E_{lp}^*$ and $E_{hump}^*$ curves, respectively. It is clearly seen that there is a third energy scale corresponding to the hump structure observed in the ARPES and tunneling spectroscopy.

We summarize the characteristic temperatures and energies derived from the spin  and charge probes in Figs.\ \ref{fig6}(e) and (f), respectively. The plotted data are summarized in Table~\ref{tab:table9}. For the characteristic temperatures, $T^*$ determined from the ($T_1T$)$^{-1}$ lies on the $T_{lp}^*$. $T^*$ determined from the neutron lies between $T_{up}^*$- and $T_{lp}^*$ curves. $T_{mK}^*$ and $T_K^*$ observed in $K_c$ lie on the $T_{hump}^*$ and $T_{up}^*$, respectively.\cite{ish98,ito98} For the characteristic energies, the half of the coherent peak energy of $B_{2g}$ ERS,\cite{ken95,sug03,hew02,gas98,tac06,ven02} half of the coherent peak energy of $B_{1g}$ ERS,\cite{pai06,blu97,cap07,gas97,mak94,sug00,kan96,ken95,sug03,hew02,gas98,tac06,ven02} and half of the two-magnon peak energy of $B_{1g}$ ERS \cite{blu97,sug03,mak94} lie on the $\Delta_c$, $E_{lp}^*$ and $E_{hump}^*$ curves, respectively.

\begin{figure*}[t]
\includegraphics[scale=0.6]{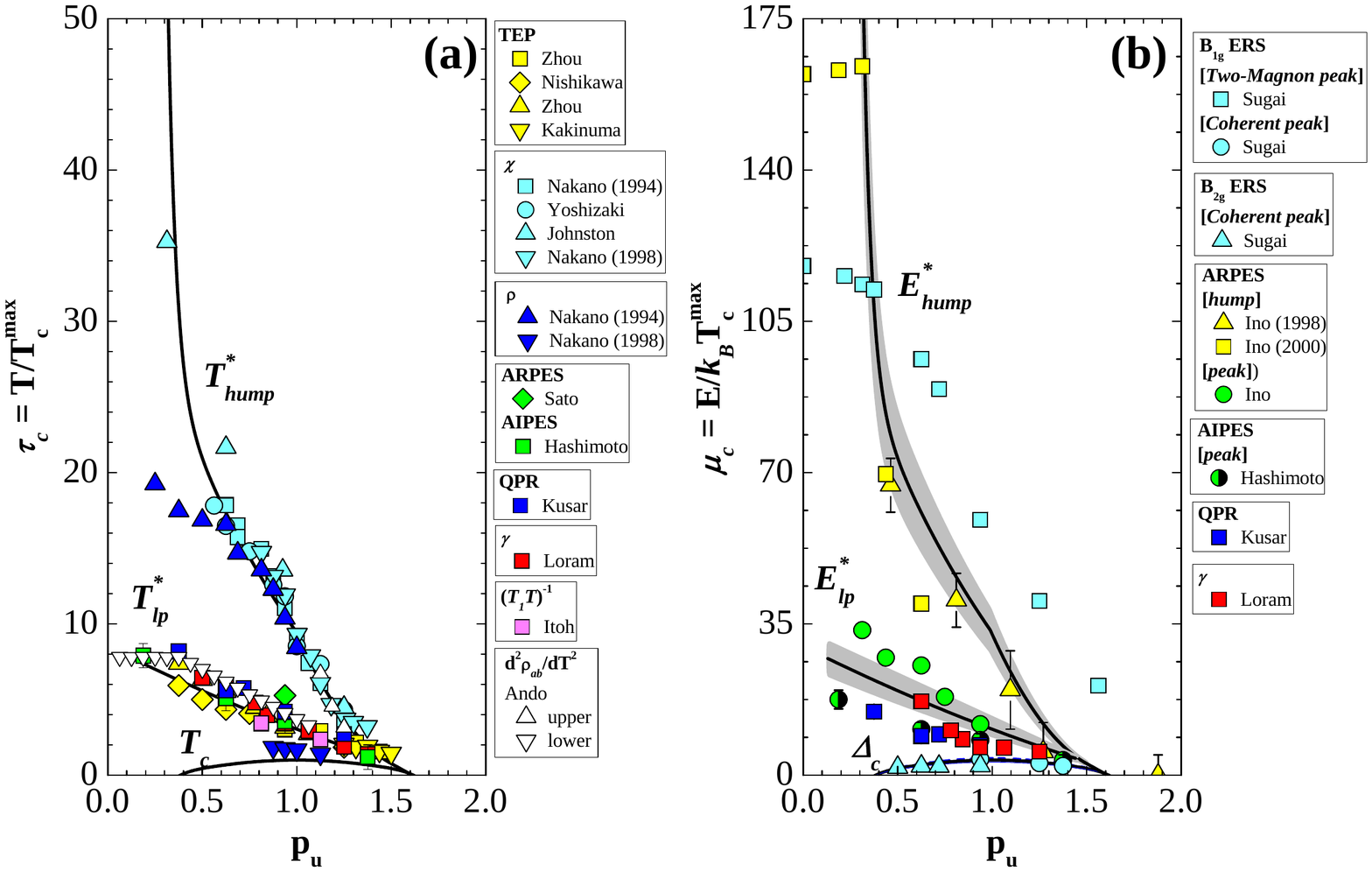}
\caption{\label{fig7} (Color online) Electronic phase diagram for the single-layer SrD-La214. The temperature- and energy-scale for the characteristic temperatures and energies are summarized in Figs.\ \ref{fig7}(a) and (b), respectively. The plotted data are summarized in Table~\ref{tab:table10}. }
\end{figure*}

From Figs.\ \ref{fig6}(a) $-$\ \ref{fig6}(f), we can conclude that the phase diagram fundamentally reproduces the $T$ vs $P_{pl}$ plot in Ref.\ \onlinecite{hon04}. Their characteristic temperatures $T^*$ lie on either the $T_{up}^*$ or the $T_{lp}^*$ curve in Ref.\ \onlinecite{hon04}. Furthermore, the third characteristic energy, the ``hump'' energy, does exist, although it is hard to detect as the corresponding characteristic temperature or $T_{hump}^*$. All four characteristic temperatures ($T_c$, $T_{lp}^*$, $T_{up}^*$, and $T_{hump}^*$) and the corresponding energies ($\Delta_c$, $E_{lp}^*$, $E_{up}^*$, and $E_{hump}^*$) do not cross each other. The four temperatures and energies tend to converge with increasing $p_u$, merge at $p_u$ $\sim$ 1.1, and finally vanish at $p_u$ $\sim$ 1.3.

Some $T^*$'s ($E^*$'s) have relatively large scattering, and are hard to group into either $T_{up}^*$ ($E_{up}^*$) -curve or $T_{lp}^*$- ($E_{lp}^*$) -curve. For example, the $E^*$ from $\kappa$ and $T^*$ from ARPES, tunneling spectroscopy and neutron scattering are scattered. These scattering may come from the differences in the characteristic time-scale and length-scale specific to different experimental probes for observing the intrinsically inhomogeneous electronic states, as discussed by Mihailovic and Kabanov.\cite{mih04} Indeed, similar to $T_{up}^*$- and $T_{lp}^*$-curves or $E_{up}^*$- and $E_{lp}^*$-curves, they all become smaller and closer in the magnitude with increasing doping in the underdoped regime. They merge into $T_c$ or $\Delta_c$-curve in the overdoped regime and universally vanish at $p_u$ = 1.3. The pseudogap, manifested either as the characteristic energy or characteristic temperature and independent of its origin, universally disappears at $p_u$ $\sim$ 1.3 together with the superconductivity. This strongly suggests that the pseudogap phase is the precursor of the superconducting phase.

 \begin{table}[b]
\caption{\label{tab:table10}The data plotted in Fig.\ \ref{fig7} for SrD-La214.}
\begin{ruledtabular}
\begin{tabular}{lll}
  \quad  Fig. & Probe & Ref(s). \\ \hline
 \quad  7(a) & TEP      & \onlinecite{kak99,nis94,zho95,coo96}  \qquad \\
\cline{2-3}
      & $\chi$   & \onlinecite{nak94,yos90,joh89,nak98} \\
\cline{2-3}
      & $\rho$   & \onlinecite{nak94,nak98} \\
\cline{2-3}
      & $d^2\rho_{ab}/dT^2$ & \onlinecite{and04} \\
\cline{2-3}
      & ARPES    & \onlinecite{sat99,ino98,ino00} \\
\cline{2-3}
      & AIPES    & \onlinecite{has07} \\
\cline{2-3}
      & specific heat & \onlinecite{lor98} \\
\cline{2-3}
      & ($T_1T$)$^{-1}$ & \onlinecite{ito04} \\
\hline
 \quad 7(b)  & ARPES    & \onlinecite{sat99,ino98,ino00} \\
\cline{2-3}
      & QPR      & \onlinecite{kus05} \\
\cline{2-3}
      & $\gamma$ & \onlinecite{lor98} \\
\cline{2-3}
      & $B_{1g}$ ERS & \onlinecite{sug03} \\
\cline{2-3}
      & $B_{2g}$ ERS & \onlinecite{sug03} \\
\end{tabular}
\end{ruledtabular}
\end{table}

\begin{figure*}[t]
\includegraphics[scale=0.6]{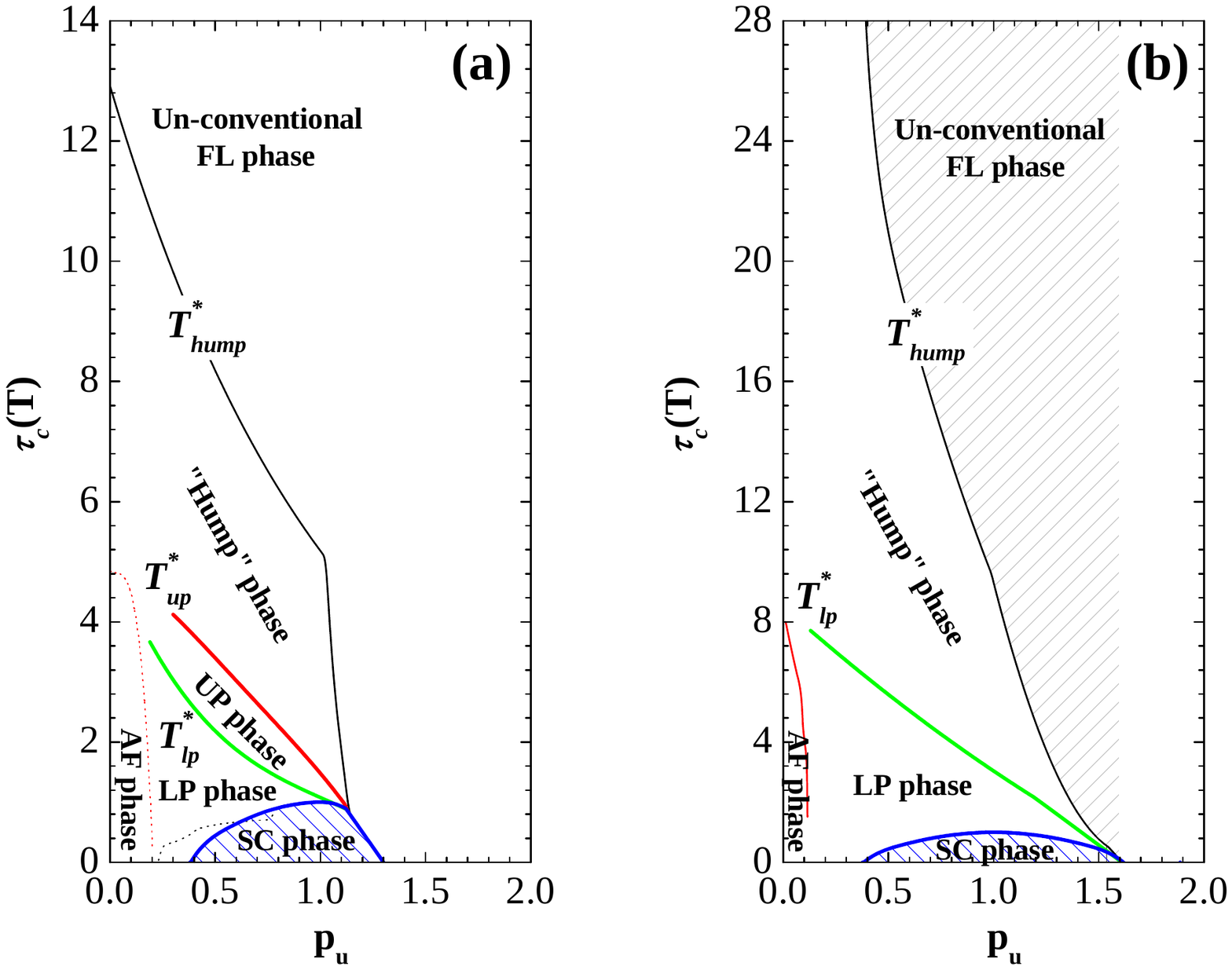}
\caption{\label{fig8} (Color online) Sketches of (a) the UEPD for HTS with $T_c^{max}$ $\sim$ 90 K and (b) the phase diagram for the SrD-La214. In Fig.\ \ref{fig8}(a), the superconducting (SC) and antiferromagnetic (AF) phases represented by the dotted lines are coming from the OD-Y123.\cite{hon07,san04} In Fig.\ \ref{fig8}(b), the AF phase for the SrD-La214 is cited from Refs.\ \onlinecite{nie98} and \onlinecite{mat02}.}
\end{figure*}

In Fig.\ \ref{fig8}(a), we present a sketch of the unified electronic phase diagram for HTSs purely based on experimental grounds. The characteristic features of the unified electronic phase diagram for single- and double-layer HTSs with $T_c^{max}$ $\sim$ 90 K are: (i) the asymmetric half-dome-shaped $T_c$-curve (SC phase), (ii) there are three characteristic temperatures, $T_{hump}^*$, $T_{up}^*$, and $T_{lp}^*$ in the underdoped region ($p_u$ $<$ 1), (iii) all three characteristic temperatures and $T_c$ come together at $p_u$ = 1.1 in the overdoped region and vanish at $p_u$ = 1.3, (iv) $T_{hump}^*$ changes into the rapid decrease at $p_u$ $\sim$ 1, (v) $T_{hump}^*$ and $T_{lp}^*$ are concave upward, while the $T_{up}^*$ is concave downward, and (vi) the electronic phase diagram on the temperature-scale can be translated into that on the energy-scale through $E$ = $zk_BT$ with $z$ = 3.5 $\pm$ 0.5. Although we use HTSs with $T_c^{max}$ $\sim$ 90 K as our model system, we should emphasize that (i) $-$ (vi) are salient features for all, except of SrD-La214 discussed in Sec. III E, HTSs. We will call this phase diagram the ``unified electronic phase diagram (UEPD)'' of HTS. Finally the $T_{lp}^*$- and $T_{up}^*$-curves tend to merge into the N\'{e}el temperature ($T_N$) curve with undoping.\cite{nie98,mat02}

\subsection{Phase diagram for the SrD-La214}

Now we discuss the HTS with symmetric $T_c$ curve, i.e., the phase diagram of SrD-La214. For the SrD-La214, the characteristic temperatures on the reduced temperature-scale and  characteristic energies on the reduced energy-scale are plotted as a function of $p_u$ in Figs.\ \ref{fig7}(a) and (b), respectively. The plotted data are summarized in Table~\ref{tab:table10}. First, noticed that the characteristic temperatures are separated into not three curves of $T_{lp}^*$, $T_{up}^*$ and $T_{hump}^*$, but two curves. Also in the energy-scale, the characteristic energies are separated into two curves. In the UEPD, the $T_{lp}^*$- or $E_{lp}^*$- curve was defined by $TEP$, the peak structure of ARPES and tunneling, $\gamma$, and $T_{hump}^*$ or $E_{hump}^*$ curve was defined by the $\rho_c$, susceptibility, the hump structure of ARPES and tunneling, and two-magnon peak of $B_{1g}$ ERS. In the SrD-La214, the lower curve of the temperature- or energy-scale is identified by $TEP$, the peak structure of ARPES and tunneling, $\gamma$, and the upper curve is identified by the $\chi$, the hump structure of ARPES and tunneling, and two-magnon peak of $B_{1g}$ ERS. Accordingly, the lower and upper curves of SrD-La214 are designated to be $T_{lp}^*$ and $T_{hump}^*$ ($E_{lp}^*$ and $E_{hump}^*$), respectively. In the SrD-La214, the $T^*$ or $E^*$ defined by the $\rho$ lie on either of the two curves, suggesting no $T_{up}^*$- or $E_{up}^*$ curve. Because, in the UEPD, the $T^*$ or $E^*$ defined by the $\rho$ and $QPR$ lied on $T_{up}^*$ or $E_{up}^*$ curve. In fact, $T_{\rho}^*$'s for the SrD-La214 with $x$ $<$ 0.16 ($p_u$ $<$ 1) lie on the $T_{hump}^*$-curve,\cite{nak94} while $T_{\rho}^*$'s for $x$ $>$ 0.14 ($p_u$ $>$ 0.875) lie on the $T_{lp}^*$ curve.\cite{nak98b} We also plot the upper and lower inflection points of $\rho_{ab}$ into the Fig.\ \ref{fig7}(a).\cite{and04} The upper and lower inflection points seem to correspond to the $T_{hump}^*$ and $T_{lp}^*$ curves, respectively. In the SrD-La214, the usual upper pseudogap temperatures identified in UEPD tend to lie on either lower pseudogap temperature or the hump temperature.

In Fig.\ \ref{fig8}(b), we present a sketch of the phase diagram for SrD-La214. The $T_c$($p_u$) follows a symmetric dome-shaped $T_c$-curve (SC phase). This is quite different from feature (i) of the UEPD. There are two characteristic temperatures, i.e., $T_{hump}^*$ and $T_{lp}^*$, in the range from the underdoped regime. This is also different from feature (ii) above. $T_{lp}^*$ of SrD-La214 seems to be a combination of $T_{up}^*$ and $T_{lp}^*$ of the UEPD. Although the $T_c$, $T_{hump}^*$ and $T_{lp}^*$ decreases with doping, there is no merging until the end point. This is also different from the feature (iii) above. $T_c$, $T_{hump}^*$, and $T_{lp}^*$ fall down to ($p_u$,$\tau_c$) = ( 1.6, 0 ) in contrast to ( 1.3, 0 ) in the UEPD. There is a slight change in curvature in the $T_{hump}^*$($p_u$) at $p_u$ $\sim$ 1. This may share the same origin as the feature (iv) above, although it is much weaker in the SrD-La214 system. Both features (v) and (vi) are similar to that of UEPD.

\subsection{Comparison between the unified electronic phase diagram and the other phase diagram}

The present UEPD is different from the phase diagrams that were proposed and discussed in Refs.\ \onlinecite{tal01,bat96,eme97} and \onlinecite{ric99}. The phase diagram in Ref.\ \onlinecite{tal01} suggests that the single $T^*$ curve crosses the dome-shaped $T_c$ curve or superconducting phase at around the optimal doping level, and fall down to $T$ = 0 at the QCP inside the superconducting phase. This phase diagram implies that there is no correlation between the pseudogap phase and high-$T_c$ phase, and therefore the pseudogap is a pure competing order. The phase diagram of Ref.\ \onlinecite{bat96} suggests that the dome-shaped $T_c$ curve intercepts the double $T^*$ curves at around the optimal doping level. The upper and lower $T^*$ curves are concave downward and upward, respectively. These phase diagram is based on the $P_{T_c}$-scale, since $T_c$ follows the superconducting dome. The phase diagram on Ref.\ \onlinecite{eme97} shows a tendency that the double $T^*$ curves merge into the asymmetric $T_c$ curve at around the slightly overdoped level and go to zero with $T_c$ at the end point. However, both $T^*$ curves are concave downward. In Refs.\ \onlinecite{tal01,bat96,eme97}, the pseudogap does not merge into the $T_N$ curve with undoping. The phase diagram discussed in Ref.\ \onlinecite{ric99} shows that the single $T^*$-curve smoothly merges into the $T_N$-curve with undoping, and smoothly merges into the asymmetric $T_c$ curve at the end point with doping. However, the single $T^*$ curve is concave downward. Thus, without alluding to the microscopic picture for the high-$T_c$ mechanism, all the previously proposed phase diagrams are different from our UEPD, except that the asymmetric $T_c$ curve in Refs.\ \onlinecite{eme97} and \onlinecite{ric99} is similar to the present half-dome shaped $T_c$ curve. 

The present UEPD clearly appears that pseudogap exists above $T_c$ for $p_u$ $<$ 1.1, while for $p_u$ $>$ 1.1 pseudogap appears at $T_c$. Even experimental data that supported a QCP inside the superconducting dome on the $P_{T_c}$-scale followed the UEPD. The phase diagram for the SrD-La214 shows that pseudogap temperatures and corresponding characteristic energies always exist above the superconducting phase, until the pseudogap disappears together with the superconducting phase at $p_u$ = 1.6. These results indicate that for all the HTSs the pseudogap phase always co-exists with the superconducting phase up to the end point and does not intersect the superconducting phase. Furthermore, the overdoped HTS with superconductivity cannot be regarded as a conventional FL phase, since there is always the pseudogap phase with superconducting phase. Both phase diagrams suggest no QCP inside the superconducting phase. Actually, it has reported that QCP may exist at around the end point of the superconducting phase when superconductivity has completely disappeared in CLBLCO,\cite{wat06} SrD-La214,\cite{ris07} OD-Tl2201,\cite{kru07} and OD-Bi2212.\cite{kru07}

The UEPD is consistent with the idea that the pseudogap phase is, if not sufficient, necessary for the high $T_c$. It also implies that at least two distinct energy scales, i.e., pseudogap and superconductivity, are required to realize high $T_c$. If we adopt a scenario that superconducting pairing is realized in the pseudogap phase and the global phase coherence occurs at $T_c$, then the smooth merging of $T^*$'s and $T_c$ in the overdoped regime suggests that cuprates become a ``more conventional'' superconductor. Because paring and phase coherence occurs at the same temperature ($T_c$). However since pseudogap still exists, it simply merges with $T_c$ and changes with $T_c$, therefore the superconducting state as well as the normal state are still ``unconventional'' up to the end point as reported in some studies.\cite{haw07,pee07} This also explains why the pseudogap phase was never observed in the overdoped regime except SrD-La214. Even in the SrD-La214, the observation of the pseudogap in the overdoped regime strongly depends on the experimental probe. For example, it is not observed in the resistivity measurements but can be clearly seen by magnetic susceptibility and TEP measurements, as shown in Fig.\ \ref{fig7}.

In the previous paper, we pointed out that the observed $T_{lp}^*$ and $T_{up}^*$ are coming from not one pseudogap, but two pseudogaps,\cite{hon04} because the temperature where the TEP has the broad peak, corresponding to $T_{lp}^*$, was different from the temperature where the TEP starts to depend on the Zn-doping, corresponding to $T_{up}^*$.\cite{hon04} However, according to the idea by Mihailovic and Kabanov,\cite{mih04} we can not completely rule out a possibility that three characteristic temperatures, including $T_{hump}^*$, are of the same physical origin. The different characteristic temperatures may come from the differences in the characteristic time-scale and length-scale specific to different experimental probes for observing the intrinsically inhomogeneous electronic states or pseudogap phase.

\section{Summary}

We have proposed a dimensionless hole-doping concentration ($p_u$), which is scaled by the optimal hole-doping concentration, for all HTSs, and construct a UEPD for almost all HTSs, except of the purely cation-doped SrD-La214. In the UEPD all experimentally observed characteristic temperatures and energies converge as $p_u$ increases in the underdoped regime, they merge together with the $T_c$ vs $p_u$ curve at $p_u$ $\sim$ 1.1 in the overdoped regime and finally goes to zero at $p_u$ $\sim$ 1.3. On the other hand, for SrD-La214, although all experimentally observed characteristic temperatures and energies converge as $p_u$ increases in the underdoped regime, they merge together with the $T_c$ vs $p_u$ curve at $p_u$ $\sim$ 1.6 where $T_c$ goes to zero. However, the detection of pseudogap becomes subtle and probe dependent for $p_u$ $>$ 1. Both the UEPD and the phase diagram of SrD-La214 clearly show that the pseudogap phase is a precursor of high $T_c$. Finally, there remains a question of why the phase diagram for SrD-La214 is different from the UEPD. The UEPD is based on the cation and oxygen co-doped HTS materials, while the SrD-La214 is the pure cation doped HTS. Although the pure oxygen-doped HTS also follows the UEPD, the phase diagram is slightly deformed by the influence of the thermally induced oxygen redistribution. Accordingly, although we can not pin down exactly why the SrD-La214 does not follow the UEPD, we speculate that the differences are coming from a combination of lattice response, such as octahedral tilt mode, to hole doping and the hard-dopant effect discussed in Ref.\ \onlinecite{lor02}. Further studies are necessary to properly address this issue.

\begin{acknowledgments}
T.H.\ would like to thank M.\ Tanimoto of Asahikawa Medical College for offering relief-time for this study. P.H.H. is supported by the State of Texas through the Texas Center for Superconductivity at the University of Houston.
\end{acknowledgments}


\end{document}